\documentclass[aps,prl,groupedaddress,twocolumn,10pt,reprint,superscriptaddress, longbibliography,floatfix]{revtex4-2}
\usepackage{epsfig,dsfont,amssymb,amsmath,amsthm,amsfonts,amsbsy,mathrsfs}
\usepackage{lipsum}
\usepackage{color}
\usepackage{graphicx}
\usepackage{amsmath}
\usepackage{amssymb}
\usepackage{subfigure}
\usepackage{float}
\usepackage{hyperref}
\usepackage{verbatim}
\usepackage{soul}
\setstcolor{red}

\def\d{{\rm d}}
\def\e{{\rm e}}

\renewcommand{\eqref}[1]{\textrm{Eq.}(\ref{#1})}

\newcommand {\be}{\begin{equation}}
\newcommand {\ee}{\end{equation}}

\begin{document}
\title{An altruistic resource-sharing mechanism for synchronization:\\ The energy-speed-accuracy tradeoff}

\author{Dongliang Zhang}
\affiliation{
Department of Physics, Tsinghua University, Beijing, 100084, China 
}
\affiliation{%
The State Key Laboratory for Artificial Microstructures and Mesoscopic Physics, School of Physics, Peking University, Beijing 100871, China}
\author{Yuansheng Cao}%
\affiliation{
Department of Physics, Tsinghua University, Beijing, 100084, China 
}%

\author{Qi Ouyang}%
\affiliation{%
Institute for Advanced Study in Physics, School of Physics, Zhejiang University, Hangzhou 310058, China
}%

\author{Yuhai Tu}%
\affiliation{IBM T. J. Watson Research Center,
Yorktown Heights, New York 10598, USA}


\begin{abstract} 
Synchronization among a group of active agents is ubiquitous in nature. Although synchronization based on direct interactions between agents described by the Kuramoto model is well understood, 
the other general mechanism based on indirect interactions among agents sharing limited resources are less known. Here, we propose a minimal thermodynamically consistent model for the altruistic resource-sharing (ARS) mechanism wherein resources are needed for individual agent to advance but a more advanced agent has a lower competence to obtain resources. We show that while differential competence in ARS mechanism provides a negative feedback leading to synchronization it also 
breaks detailed balance and thus requires additional energy dissipation besides the cost of driving individual agents. By solving the model analytically, our study reveals a general tradeoff relation between the total energy dissipation rate and the two key performance measures of the system: average speed and synchronization accuracy. For a fixed dissipation rate, there is a distinct speed-accuracy Pareto front traversed by the scarcity of resources: scarcer resources lead to slower speed but more accurate synchronization. Increasing energy dissipation eases this tradeoff by pushing the speed-accuracy Pareto front outwards. The connections of our work to realistic biological systems such as the KaiABC system in cyanobacterial circadian clock and other theoretical results based on thermodynamic uncertainty relation are also discussed. 

\end{abstract}
\maketitle

{\bf Introduction}. Collective behaviors are ubiquitous in nature across different scales from coordinated motion of molecular motors to bacterial swarming to bird flocking ~\cite{marchetti2013hydrodynamics,bialek2012statistical,toner1998flocks,elgeti2015physics}. Despite their diversity, these systems all share a common feature: they operate out-of-equilibrium and cost free energy, which are used to counter the effects of inevitable internal and external noise in order to sustain the collective order\cite{yu2022energy,sync_energy,acebron2005kuramoto,tailleur2022active}.  

Synchronization of molecular systems~\cite{Wolde2007, Sasai2010, Wolde2017} is a collective phenomenon critical for maintaining biological functions. 
Mechanistically, synchronization can be achieved by either direct or indirect interactions between individual molecular systems. In the KaiABC system for the circadian clock of cyanobacteria, both the direct and indirect interaction mechanisms may be at work. The direct interactions are carried out by monomer shuffling between the KaiC hexamers \cite{Kondo2006, emberly2006, Ito2007, Johnson2007, Sasai2007, Sasai2008, Sasai2010}; and the indirect interactions among KaiC hexamers are mediated by the shared pool of KaiA molecules, which are the key regulators for phosphorylation reactions that generate the circadian rhythm for KaiC hexamers~\cite{Wolde2007,rust2007,ma2012,Rust2014,Wolde2017,suri2021Kai,mori2018}. 


In a previous work~\cite{sync_energy}, we studied thermodynamics of the direct interaction mechanism for synchronization based on monomer-shuffling of the KaiC hexamers, and discovered a tradeoff relation between the energy cost and synchronization accuracy. In this paper, we develop a thermodynamically consistent model to study the energy cost of the indirect interaction mechanism for synchronization inspired by KaiA differential binding in the KaiABC system (see Fig.~S1 in the Supplementary Information (SI) for an illustration of the two mechanisms). We find that the same energy-speed-accuracy relation holds as in the direct interaction mechanism suggesting a universal relation between energy cost and synchronization performance independent of detailed mechanisms.  

{\bf The altruistic resource-sharing model.}
We consider $N$ agents each with a state variable $x$, and $M$ resource molecules (activators). When bound with the activator, an agent can advance through a biased random walk process: the rate to step forward is $k$, and the rate to step backward is $\gamma k$, where $\gamma=\e^{-e_p\Delta x}<1$ with 
$e_p\equiv-\ln\gamma/\Delta x>0$ representing the driving energy for the biased random walk per unit ``length" (step).
When an agent is not bound with an activator, it is purely diffusive with a slow stepping rate $k'(\ll k)$. Here, we take $k'=0$ for simplicity. 

The binding and unbinding rate for the activator 
is $q$ and $q\e^{E(x)}$, respectively, where $E(x)$ is the binding energy for the agent with state variable $x$. In the altruistic resource-sharing model, $E(x)$ increases with $x$, which creates an effective negative feedback to prevent leading (``rich") agents from getting more resources. 
For simplicity, we use a linear form for the binding energy: $E(x)=\alpha x$, where $\alpha>0$ is a parameter describing the binding energy difference per unit length.  A larger $\alpha$ indicates a larger affinity difference and stronger negative feedback. See SI for a detailed discussion for having a nonlinear $E(x)$. 

\begin{figure}
\begin{minipage}{\linewidth}
\centering
\includegraphics[width=0.99\linewidth]{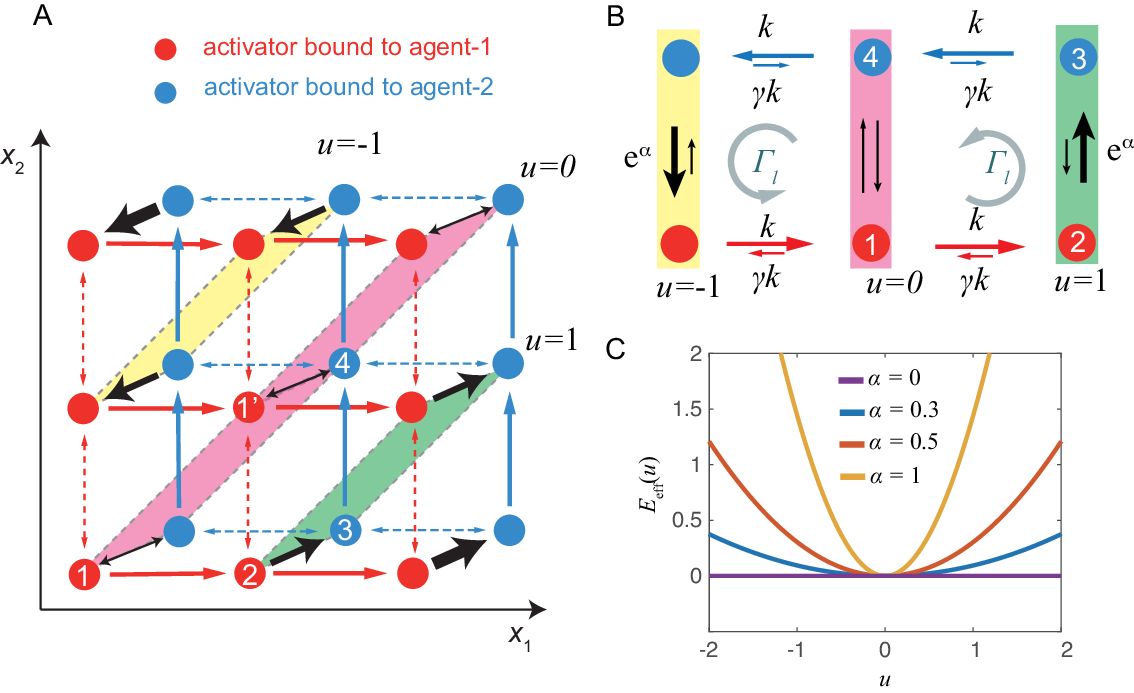}
\end{minipage}
\caption{Illustration of the simplest model ($M=1$, $N=2$). 
(A) Each state (dot) is defined by coordinates of the two agents $(x_1,x_2)$ and its color represents which agent the activator is bound to (red: agent 1; blue: agent 2). The red-states (blue-states) are biased to move right (up). Altruistic differential binding ($\alpha>0$) leads to the preferred transitions (black arrows) from a red-state to the blue-state when $u(\equiv x_1-x_2)>0$ or vise versa when $u<0$. (B) Projection onto the $(u,``color")$ plane to illustrate the dissipative cycle formed by directed motions (horizontal transitions) and differential binding (vertical transitions). 
(C) The effective potential $E_{eff}(u)$ arises with finite $\alpha>0$. 
}
\label{fig:model}
\end{figure}

{\bf The simplest case}. To gain intuition about the dynamics and energetics of the altruistic resource-sharing (ARS) model, we first investigate the simplest case with two agents and one activator: $N=2$, $M=1$. Furthermore, we consider the situation ($e^{E(x)}\ll 1$) 
where there is no unbound (free) activator. The activator can be bound either to agent-1 or agent-2 as represented by the red and blue dotes in Fig.~\ref{fig:model}A, respectively.  

As shown in Fig.~\ref{fig:model}A, there are two types of transitions (arrowed lines) in the system. The horizontal (red) and vertical (blue) arrowed lines correspond to the processive transitions (directed motions) of the individual agents. The out-of-plane (black) lines correspond to the switching of the activator from one agent to another. Synchronization can be evaluated by the difference between the two agents: $u\equiv x_1-x_2$. The synchronized plane with $u=0$ and the two adjacent less synchronized plane with $u=\pm 1$ are shown in Fig.~\ref{fig:model}A. Starting with a red state (1) on the $u=0$ plane (the pink plane in Fig.~\ref{fig:model}A), the directed motion of agent-1 will lead to a less synchronized state (2) on the $u=1$ plane (the green plane in Fig.~\ref{fig:model}A). However, if $\alpha>0$, the activator will leave agent-1 and bind with agent-2 leading to the blue state (3) on the $u=1$ plane. The subsequent processive motion of agent-2 will bring the system back to a state (4) on the synchronized ($u=0$) plane again.

By going to the co-moving frame, all processive transitions can be projected onto the $u$-direction. As shown in Fig.~\ref{fig:model}B, the combination of processive motion and differential binding ($\alpha>0$) forms the dissipative flux cycle quantified by the ratio between the products of rates in counter-clock-wise and clock-wise directions: 
\begin{equation}\label{eq:rate_ratio}
\Gamma_l\equiv \frac{\prod_{ccw}\text{(rates)}}{\prod_{cw}\text{(rates)}}=\frac{e^{\alpha\Delta x}}{\gamma^2}=\e^{(2e_p+\alpha)\Delta x},
\end{equation}
which shows that the system operates out of equilibrium ($\Gamma_l\ne 1$) due to both directed motion ($e_p>0$) and differential affinity ($\alpha>0$). 

Quantitatively, we can solve the Fokker-Planck equation for the joint probability density $P(x_1,x_2,t)$ in the continuum limit of $\Delta x \to 0$~\footnote{we did rescaling $k(\Delta x)^2\to k$ in the continuum limit}:
\begin{equation}
\frac{\partial P}{\partial t}=
-\sum_{i=1}^2\frac{\partial}{\partial x_i}\left(ke_p p_iP-\frac{\partial}{\partial x_i}kp_iP
\right),
\end{equation}
where fast binding is assumed ($q\gg k$) and the probability $p_i$ for the $i$-th agent to be bound with the activator is given by: 
$p_i={\e^{-E(x_i)}}/(\e^{-E(x_1)}+\e^{-E(x_2)})$. 
We show that the average position $\bar{x}\equiv(x_1+x_2)/2$ has a mean velocity $v=ke_p/2$ and the steady state distribution $P_s(u)$ of the difference $u$ is given by (see SI for detailed derivation):
\begin{equation}
   P_s(u)=Z_2^{-1}\exp{(-\frac{2e_p}{\alpha}\ln\cosh{\frac{\alpha u}2})},
\end{equation}
where $Z_2$ is the normalization constant. From $P_s(u)$, we can define an effective potential $E_{eff}$: 
$E_{eff}(u)\equiv -\ln(P_s(u))=2\frac{e_p}{\alpha}\ln\cosh{\frac{\alpha u}2}$. As shown in Fig.~\ref{fig:model}C, $E_{eff}(u)$ is flat for $\alpha=0$, and it has a minimum at $u=0$ for $\alpha>0$ and thus stabilizes the synchronized state. 

{\bf The thermodynamic limit.}
In the thermodynamic limit $N\to\infty, M\to\infty$ with a finite $M/N\equiv m_t$, 
the system is described by the probability density function $\rho(x,t)$, which satisfies the Fokker-Planck equation:
\begin{equation}\label{mean-field}
\frac{\partial \rho(x,t)}{\partial t}=
-k\frac{\partial}{\partial x}\left(e_pp(x)\rho-\frac{\partial}{\partial x}p(x)\rho
\right),
\end{equation}
where $p(x)$ is the activator-occupancy function. 
Under the fast binding assumption, we have $p(x)=g/[\e^{\alpha(x-x_g)}+1]$ 
with $g(\ge 1)$ the maximum activator occupancy number and $x_g$ a time-dependent parameter, which is determined by conservation of activators:
\begin{equation}
\int p(x)\rho(x,t)\d x=\int\frac{g}{\e^{\alpha(x-x_g)}+1}\rho(x,t)\d x=m_t,
\label{p_eq}
\end{equation}
where we have neglected the number of free activators (see SI for details).


Remarkably, Eqs.~\ref{mean-field}\&\ref{p_eq} can be solved analytically to obtain a steady-state travelling wave solution $\rho(x,t)=\rho_s(u)$ where $u=x-vt$ is the relative position and $v=ke_pm_t$ is the average drift speed of the population. The expression of $\rho_s(u)$ is (see SI for detailed derivation):
\begin{equation}\label{eq:rho_g}
\rho_{s}(u)=\frac1Z(e^{\alpha u}+1)\exp[e_p(1-A_t) u-\frac{A_te_p}{\alpha}e^{\alpha u}],
\end{equation}
where $A_t\equiv m_t/g>0$ describes the resource abundance relative to demand and $Z$ is the normalization constant.

\begin{figure}
\begin{minipage}{\linewidth}
\centering
\includegraphics[width=0.99\linewidth]{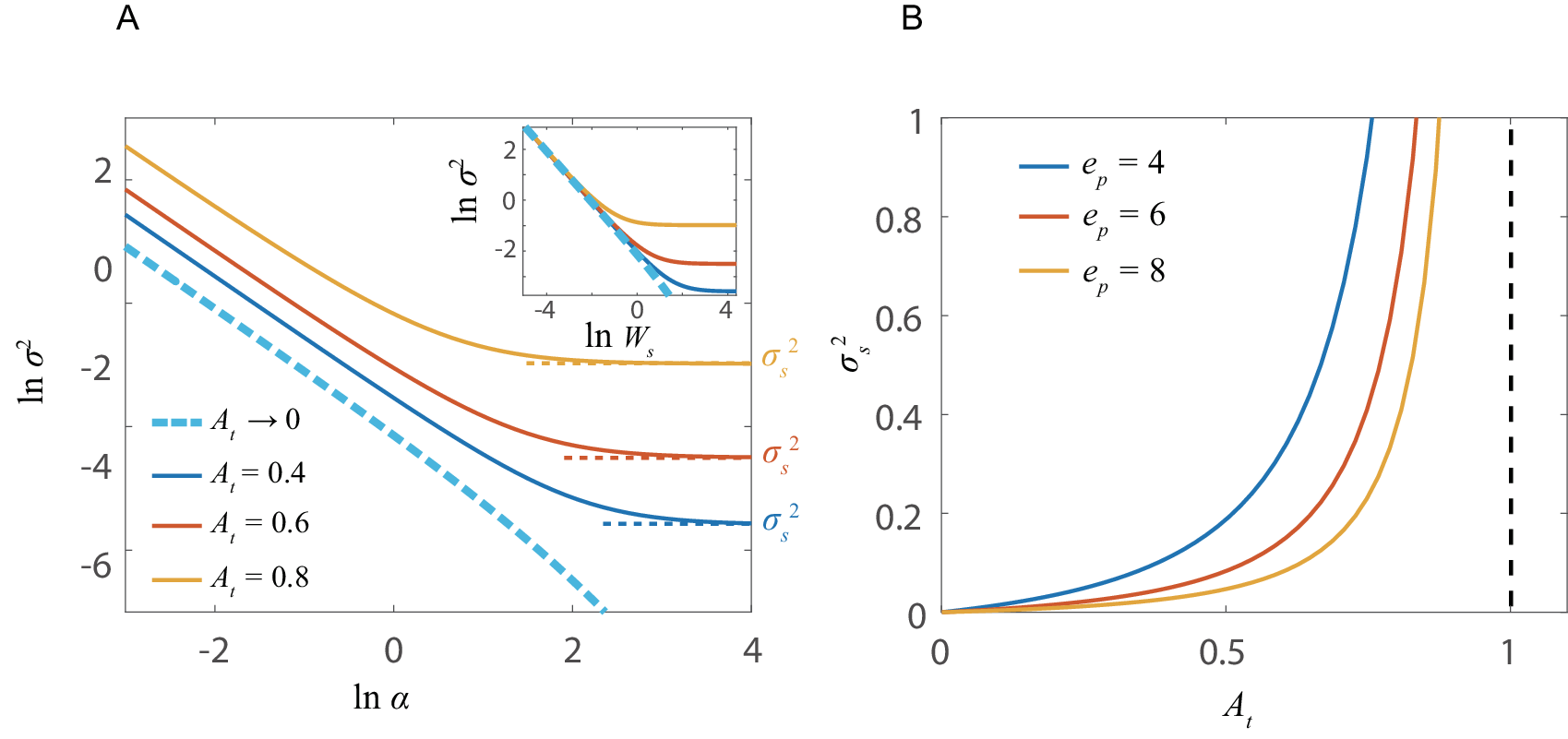}
\end{minipage}
\caption{Dependence of synchronization on key parameters. 
(A) The dependence of $\sigma^2$ on $\alpha$ with fixed $e_p=8$ and different $A_t$. $\sigma$ decreases with $\alpha$ and saturates to a finite value $\sigma_s$ for large value of $\alpha$. Inset: The dependence of $\sigma^2$ on dissipation $W_s$ collapses onto the same curve for different values of $A_t$ before saturation.   
(B) The dependence of the saturation variance $\sigma^2_s$ on $A_t$ for different $e_p$. $\sigma_s\to0$ when $A_t\to0$ and increases with $A_t$. $\sigma_s\to\infty$ when $A_t\to1$.
}\label{fig:numeric}
\end{figure}

We use variance of $u$ to characterize the synchronization performance: $\sigma^2\equiv\int(u-\bar{u})^2\rho_s(u)\d u$ with $\bar{u}\equiv\int u\rho_s(u)\d u$. 
Smaller $\sigma$ indicates better synchronization. Direct numerical simulations show that $\sigma^2$ decreases with $\alpha$ (Fig.\ref{fig:numeric}A) but always saturates to a finite value $\sigma_s^2$ that increases with $A_t$ (Fig.\ref{fig:numeric}B). This behavior can be understood from the analytical expression of the steady-state distribution $\rho_s(u)$ given in Eq.~\ref{eq:rho_g}. 

In the weak sharing limit $\alpha\ll 1$, by expanding the effective potential $E_{eff}\equiv -\ln(\rho_s)$ near its minimum, we have 
a leading order estimation for $\sigma^2$: 
\begin{equation}
    \sigma^2\approx \frac1{\alpha e_p(1-A_t)},\;\;\;\; \alpha\ll 1
    \label{eq:alpha_s}
\end{equation} which clearly shows that synchronization improves with increasing $\alpha$ and $e_p$ in agreement with direct simulation results shown in Fig.~\ref{fig:numeric}A. 

In the strong sharing limit $\alpha\gg 1$, $\rho_s$ becomes the sum of a $\delta$-function distribution and an exponential distribution (see SI for details):
\begin{equation}\label{eq:rho_infty}
\rho_{s,\infty}(u)=C_1\delta(u)+C_2\rho_e(u)H(-u), 
\end{equation}
where $H$ is the Heaviside step function and $\rho_e(u)=\lambda \e^{\lambda u}$ with $\lambda=e_p(1-A_t)$. $C_1=1-A_t$ and $C_2=A_t$ are the weights of the two distributions. By using \eqref{eq:rho_infty}, we obtain an analytical expression of the saturation variance $\sigma_s$: 
\begin{equation}
\sigma^2(\alpha\rightarrow \infty)\equiv \sigma_s^2=\frac{(2-A_t)A_t}{e_p^2(1-A_t)^2},
 \label{eq:alpha_l}
\end{equation}
which diverges as $A_t=1$ and becomes zero as $A_t=0$; and it also decreases with the processive driving force $e_p$ all consistent with numerical results shown in Fig.~\ref{fig:numeric}B.

Besides the importance of differential binding ($\alpha>0$), our results also reveal that synchronization depends critically on the relative resource abundance ($A_t$). Based on Eq.~\ref{eq:alpha_s}, the variance $\sigma^2$ diverges when $A_t\rightarrow 1$, which means that the ARS mechanism only works when the resource is relatively scarce ($A_t<1$) and synchronization improves as $A_t$ decreases. In the large $\alpha$ limit (Eq.~\ref{eq:rho_infty}), $C_1$ represents the fraction of agents that are perfectly synchronized (delta-function distribution of $u$). The expression for $C_1=1-A_t$ reveals the surprising finding that the system is only partially synchronized ($C_1<1$) even in the limit of infinite sharing ($\alpha\rightarrow \infty$). Indeed, Eq.~\ref{eq:alpha_l} shows that the system only becomes perfectly synchronized in the limit of both $\alpha\rightarrow \infty$ and $A_t\rightarrow 0$.   

{\bf The energy-speed-accuracy (ESA) relation.} Under the fast binding approximation, the free energy dissipation rate per agent can be computed by following previous work~\cite{tome2010entropy,zhang2021entropy}:
\begin{equation}
    \dot{W}\approx\int\frac{J^2}{kp\rho}\d\vec{x}=m_t ke_p^2
+k\int\frac1{p\rho}\left(\frac{\partial(p\rho)}{\partial x}\right)^2\d\vec{x},
\label{eq:diss}
\end{equation}
where $J\equiv ke_pp\rho-k\partial_{x}(p\rho)$ is the processive probability flux. 
The first term in Eq.~\ref{eq:diss}, denoted by $\dot{W}_0=m_t k e_p^2=ve_p$ with $v=m_t k e_p$ the average speed, is the dissipation for driving the processive dynamics of individual agent. The second term, denoted by $\dot{W}_s$, is the dissipation for driving differential binding.
Using expressions for $p(x)$ and $\rho_s(x)$, we have $\dot{W}_s=v(1-A_t)\alpha$ (see SI for details). Put together, the total dissipation can be written as:
\begin{equation}\label{eq:Ws}
\dot{W}=\dot{W}_0+\dot{W}_s=ve_p + v(1-A_t)\alpha.
\end{equation}

By using Eqs.~\ref{eq:alpha_s}\&\ref{eq:alpha_l}\&\ref{eq:Ws}, we can quantitatively relate the free energy dissipation and the synchronization error given by $\sigma^2$:
\begin{equation}\label{sigma_w}
\sigma^2
\approx
\left\{
\begin{aligned}
   &\frac1{W_sW_0}, \;\;\; &W_s\ll W_{c},  \\
   &\frac{(2-A_t)A_t}{W_0^2(1-A_t)^2},\;\;\; &W_s\gg W_{c},
\end{aligned}
\right.
\end{equation}
where $W_s\equiv \dot{W}_s/v=(1-A_t)\alpha$ and $W_0\equiv \dot{W}_0/v=e_p$ are the dissipation per unit length for driving synchronization and processive motion, respectively, and $W_c\equiv \frac{(1-A_t)^2}{A_t (2-A_t)} W_0$ is a crossover dissipation.

Eq.~\ref{sigma_w} clearly shows the tradeoff between synchronization error and energy dissipation. The inverse $\sigma^2$-$W_s$ dependence for different values of $A_t$ collapse onto the same curve before they saturate beyond the crossover dissipation $W_c$, which decreases with $A_t$ and increases with $W_0$ (see inset of Fig.~\ref{fig:numeric}A). While synchronization is more accurate for lower resource abundance (smaller $A_t$) and stronger directed motion (larger $e_p$)
; the speed $v(=e_p A_t)$, which is the other important measure of the system's performance, increases with both $e_p$ and $A_t$. 


\begin{figure}
\begin{minipage}{\linewidth}
\centering
\includegraphics[width=0.99\linewidth]{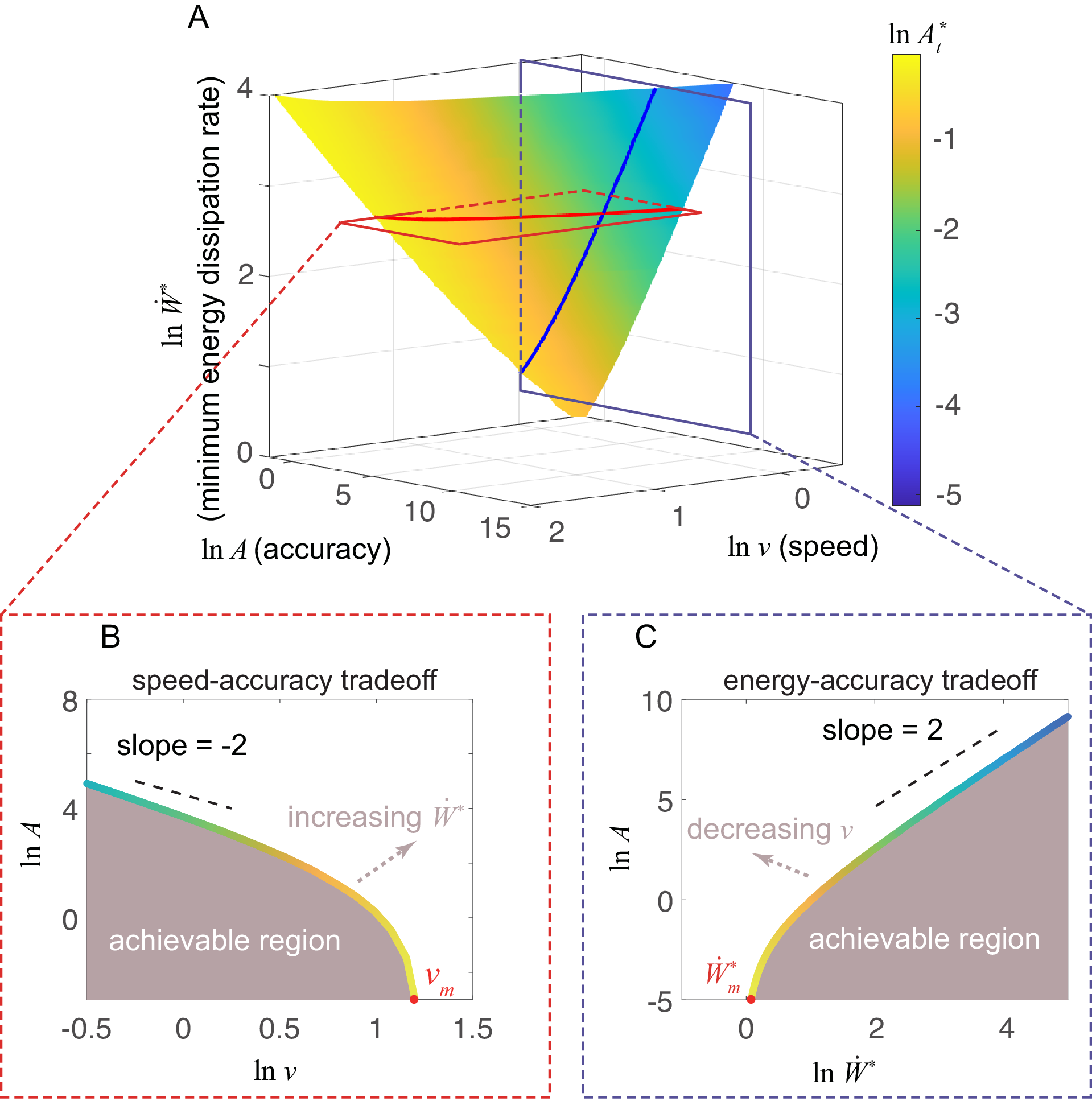}
\end{minipage}
\caption{
Energy-speed-accuracy tradeoff.
(A) The Pareto surface, i.e., the minimum required total energy dissipation rate $\dot{W}^*$ to achieve a certain speed-accuracy $(v,A)$. 
The color on the surface represents the optimal $A_t^*$ to achieve the minimum $\dot{W}^*$. 
(B) The tradeoff between speed ($v$) and synchronization accuracy ($A$) with a fixed dissipation. The $(v-A)$ Pareto front shifts outwards with increasing dissipation rate $\dot{W}^*$.   
(C) The tradeoff between accuracy $A$ and dissipation rate $\dot{W}^*$  with a fixed speed $v$.  
}\label{fig:ESAtradeoff}
\end{figure}


By sweeping through the parameter space $(A_t,e_p,\alpha)$, we studied the minimum energy dissipation rate $\dot{W}^*$ for achieving a given system-level performance characterized by the synchronization accuracy $A\equiv \sigma^{-2}$ and average speed $v$. As shown in Fig.~\ref{fig:ESAtradeoff}A, the 2D ``Pareto" surface $\dot{W}^*(v,A)$ represents the energy-speed-accuracy(ESA) relationship: any performance $(v,A)$ inside the Pareto surface is achievable with a minimum energy dissipation rate $\dot{W}\ge\dot{W}^*(v,A)$; performance outside the Pareto surface is forbidden. Each point $(\dot{W}^*, v, A)$ on the Pareto surface is achieved by an ``optimal" set of parameters $(A^*_t,e^*_p,\alpha^*)$. In Fig.~\ref{fig:ESAtradeoff}, the optimal $A^*_t$ is represented by the color on the Pareto surface~\footnote{$e_p^*$ and $\alpha^*$ can be determined from $A_t^*$ and the targeted performance $(v,A)$}. 

To understand the ESA tradeoff, we dissect the ESA Pareto surface along different directions. In Fig.~\ref{fig:ESAtradeoff}B, we cut the Pareto surface at a fixed minimum dissipation, which shows a Pareto front in the performance space $(v,A)$ that clearly demonstrates the accuracy-speed tradeoff modulated by the choice of $A^*_t$ represented by the color along the Pareto front. As $A^*_t$ changes along the Pareto front, the system goes from higher accuracy lower speed at smaller $A_t^*$ to lower accuracy higher speed at larger $A_t^*$. For a fixed $\dot{W}^*$, there is a maximum speed $v_m=(kg\dot{W}^*)^{1/2}$; for $v\ll v_m$, we have $A\propto v^{-2}$. As we increase the minimum dissipation $\dot{W}^*$, the Pareto front moves outward extending the achievable performance region. We can also cut the Pareto surface at a fixed speed $v$ and plot the accuracy versus the minimum dissipation (Fig.~\ref{fig:ESAtradeoff}C), which shows that there is a minimum dissipation $\dot{W}_m=v^2/(kg)$ needed for achieving a finite accuracy. For $\dot{W}^*\gg W_m$, we have $A\propto \dot{W}^{*^2}$, and the higher accuracy is driven by a decreasing $A^*_t$ (color of the line). 

The same dependence of synchronization error on energy dissipation as Eq.~\ref{sigma_w} is found in the other general synchronization mechanism based on direct interactions between agents~\cite{sync_energy} (see SI for a detailed derivation). Together, our results suggest that the ESA relation is universal in synchronization of molecular systems.





{\bf Comparison with the thermodynamic uncertainty relation (TUR)}:
A new class of inequalities called the thermodynamic uncertainty relation (TUR), which relates entropy production with the mean and variance of an observed flux (current), have been used to set a lower bound for the dissipation
~\cite{barato2015thermodynamic, TUR, lee2018thermodynamic,dieball2023direct}. Following previous study on TUR for synchronized oscillators \cite{lee2018thermodynamic}, we focus on an observable $l_T$, which corresponds to the distance an agent travels during time interval $T$:
$
l_T=\int_{-\infty}^{+\infty} j_{x,T}\d x,
$
where $j_{x,T}$ is the net number of transitions from state $x$ to $x+\d x$ within time $T$. 

From the steady state solution of our system, the average $l_T$ is: $\langle l_T \rangle=vT$, and the variance $\sigma_l^2$ is given by:
\begin{align}    
    \sigma_l^2=\frac{T}{N}\int 2kp(x)\rho(x)\d x+\sigma^2
    =\frac{2vT}{Ne_p}+\sigma^2,
\end{align}
where the first and the second terms represent the variance of mean position of all the $N$ agents and the relative position of an agent to the mean, respectively (see SI for details). 
From TUR, we can obtain a lower bound ($\dot{W}_{TUR}$) for the total dissipation rate  per agent: 
\begin{equation}
   \dot{W}_{TUR}\equiv \frac{2\langle l_T\rangle^2}{NT\sigma^2_l}=\frac{\dot{W}_0}{1+\frac{e_p\sigma^2 N}{2vT}}.
   \label{wtur}
\end{equation}

From Eq.~\ref{eq:Ws}, the exact value of dissipation is given by: $\dot{W}=\dot{W}_0(1+(1-A_t)\alpha/e_p)$. It is clear that the TUR bound $\dot{W}_{TUR}$ obtained from statistics of a local flux does not provide an accurate estimate for the dissipation of the whole system $\dot{W}$. Most evidently, while the minimum dissipation $\dot{W}$ is the baseline dissipation $\dot{W}_0=v e_p$ without synchronization ($\alpha=0$); the maximum value of $\dot{W}_{TUR}$ is $\dot{W}_0$ reached only at $\alpha\to\infty$ (see Fig.~S3 in SI for details). 
%
From Eq.~\ref{wtur}, the equality $\dot{W}_{TUR}=\dot{W}$ only occurs for infinite observation time $T\to\infty$ and in the trivial case with $\alpha=0$ when $\rho(x)$ is an uniform distribution and there is no synchronization~\cite{dieball2023direct}. Our result reflects the limitation of using TUR for specific individual observables to obtain accurate estimate of the dissipation in the whole system.

{\bf Discussion.} In this work, by introducing a minimal model to study the altruistic resource-sharing (ARS) mechanism for synchronization with indirect agent-agent interaction, we showed that the ARS mechanism necessarily costs energy to facilitate differential binding critical for synchronization. The energy cost, the collective speed of the agents, and the accuracy of synchronization satisfy the same energy-speed-accuracy (ESA) tradeoff relation found in the direct pair-wise interacting (PI) mechanism studied previously~\cite{sync_energy}, which suggest that the ESA relation for synchronization is independent of details of the underlying mechanisms. Remarkably, sensory adaptation systems~\cite{Tu2018Adaptation} also exhibit an universal ESA relation independent of whether the underlying control mechanism is negative feedback~\cite{lan2012energy} or incoherent feed-forward~\cite{lan2013cost}. The ESA relation reveals that increasing energy dissipation can ease the tradeoff between key performance measures, e.g., speed and accuracy as shown in Fig.~\ref{fig:ESAtradeoff}. This general functional role of energy dissipation also applies to other nonequilibrium biological systems including sensory adaptation~\cite{lan2012energy,Sartori2015Free}, biochemical oscillation~\cite{Cao2015Free-energy, Fei2018Design}, and signal transduction~\cite{Hathcock2023Nonequilibrium,Hathcock2024Time-reversal}.

Aside from their common ESA relation, it is important to highlight two key differences between the ARS and PI mechanisms for synchronization. First, while the PI mechanism depends on direct interactions between agents, the ARS mechanism requires additional resource particles (activators) to mediate the indirect interactions between agents. In particular, differential binding of the agents to the activator ($\alpha>0$) mediates an effective (indirect) interaction among agents that favors synchronization. 
Second, for the PI mechanism, 
the coupling between agents is introduced by additional chemical reactions with rates and strength that are independent of the processive dynamics of individual agents. 
However, for the ARS mechanism, 
the differential affinity of the resource ($\alpha$) directly affects the processive dynamics of individual agents. 
As a result, the synchronization performance in ARS depends not only on $\alpha$ but also explicitly on the processivity of the agents characterized by $e_p$ as shown by the expression of $\sigma^2$ given in Eqs.~\ref{eq:alpha_s}\&\ref{eq:alpha_l}.

The key ingredient in the ARS mechanism is differential binding. In the 
KaiABC system, the nucleotide exchange factor KaiA, which facilitates the successive phosphorylation of KaiC hexamer, has a higher affinity to the KaiC hexamers with a lower phosphorylation level \cite{Rust2014, KaiA_affinity}. 
This suggests that the ARS mechanism may be responsible for synchronization of the KaiC hexamers\cite{takigawa2006predicting,clodong2007functioning,miyoshi2007,Wolde2007,axmann2007minimal}. Another evidence for the ARS mechanism is that the total KaiA concentration ($A_t$) plays an important role in controlling synchronization: a relatively low $A_t$ is essential for synchronization, which is consistent with experiments that show the disappearance of coherent oscillation for high levels of KaiA\cite{nakajima2010vitro,chavan2021reconstitution}. 

In addition to the differential binding of KaiA, previous studies~\cite{emberly2006,sync_energy} have shown that monomer shuffling between KaiC hexamers—an experimentally observed phenomenon~\cite{Kondo2006, Ito2007, Johnson2007, Sasai2007, Sasai2008, Sasai2010}—can also contribute to synchronization through the PI mechanism. Indeed, existing experiments suggest that both the differential binding and monomer-shuffling mechanisms may contribute to synchronization in the KaiABC system. Moreover, it is crucial to note that the differential binding mechanism, which relies on a monotonically increasing $E(x)$, operates exclusively during the phosphorylation phase. Interestingly, monomer-shuffling has been observed to occur more frequently during the dephosphorylation phase~\cite{Ito2007}. This presents an intriguing opportunity for future research to explore how these two mechanisms might act synergistically to achieve precise and efficient synchronization in the KaiABC system. 

Resource sharing is an ubiquitous phenomenon in living systems across all scales. For example, resource sharing is critical in driving cooperative behaviors in ecology and evolution~\cite{Bonsall2012Altruism,Kreider2022resource}. Here, we showed how altruistic resource sharing (ARS) can lead to synchronization in molecular systems. The two key ingredients of ARS, the limited resource and differential affinity, are also present in various other
molecular systems, such as bacterial DNA replication initiation \cite{mott2007dna,fu2023bacterial} and combinatorial signal processing in the BMP pathway~\cite{Antebi2017combinatorial, su2022ligand, klumpe2022context, klumpe2023computational}. It would be interesting to test if similar cost-performance tradeoff relation as the ESA relation for synchronization could be found in these systems.

%



{\bf Acknowledgement.}
The work by Y. T. is supported by a NIH Grant (R35GM137134). The work by Q. O. is supported by the Starry Night Science
Fund of Zhejiang University Shanghai Institute for Advanced Study. 
The work by D. Z. and Y. C. are supported by the National Natural Science Foundation of China (Grant No. 12090054 and Grant No. 12374213). Y. T. would like to thank the Center 
for Computational Biology at the Flatiron Institute for 
hospitality while a portion of this work was carried out. 
D. Z. would like to thank Shiling Liang and other colleagues in his current institute, Max Planck Institute for the Physics of Complex Systems, for discussion and support.

\bibliography{Sync_diff_ref.bib}

\begin{thebibliography}{53}%
\makeatletter
\providecommand \@ifxundefined [1]{%
 \@ifx{#1\undefined}
}%
\providecommand \@ifnum [1]{%
 \ifnum #1\expandafter \@firstoftwo
 \else \expandafter \@secondoftwo
 \fi
}%
\providecommand \@ifx [1]{%
 \ifx #1\expandafter \@firstoftwo
 \else \expandafter \@secondoftwo
 \fi
}%
\providecommand \natexlab [1]{#1}%
\providecommand \enquote  [1]{``#1''}%
\providecommand \bibnamefont  [1]{#1}%
\providecommand \bibfnamefont [1]{#1}%
\providecommand \citenamefont [1]{#1}%
\providecommand \href@noop [0]{\@secondoftwo}%
\providecommand \href [0]{\begingroup \@sanitize@url \@href}%
\providecommand \@href[1]{\@@startlink{#1}\@@href}%
\providecommand \@@href[1]{\endgroup#1\@@endlink}%
\providecommand \@sanitize@url [0]{\catcode `\\12\catcode `\$12\catcode
  `\&12\catcode `\#12\catcode `\^12\catcode `\_12\catcode `\%12\relax}%
\providecommand \@@startlink[1]{}%
\providecommand \@@endlink[0]{}%
\providecommand \url  [0]{\begingroup\@sanitize@url \@url }%
\providecommand \@url [1]{\endgroup\@href {#1}{\urlprefix }}%
\providecommand \urlprefix  [0]{URL }%
\providecommand \Eprint [0]{\href }%
\providecommand \doibase [0]{http://dx.doi.org/}%
\providecommand \selectlanguage [0]{\@gobble}%
\providecommand \bibinfo  [0]{\@secondoftwo}%
\providecommand \bibfield  [0]{\@secondoftwo}%
\providecommand \translation [1]{[#1]}%
\providecommand \BibitemOpen [0]{}%
\providecommand \bibitemStop [0]{}%
\providecommand \bibitemNoStop [0]{.\EOS\space}%
\providecommand \EOS [0]{\spacefactor3000\relax}%
\providecommand \BibitemShut  [1]{\csname bibitem#1\endcsname}%
\let\auto@bib@innerbib\@empty
\bibitem [{\citenamefont {Marchetti}\ \emph {et~al.}(2013)\citenamefont
  {Marchetti}, \citenamefont {Joanny}, \citenamefont {Ramaswamy}, \citenamefont
  {Liverpool}, \citenamefont {Prost}, \citenamefont {Rao},\ and\ \citenamefont
  {Simha}}]{marchetti2013hydrodynamics}%
  \BibitemOpen
  \bibfield  {author} {\bibinfo {author} {\bibfnamefont {M.~C.}\ \bibnamefont
  {Marchetti}}, \bibinfo {author} {\bibfnamefont {J.-F.}\ \bibnamefont
  {Joanny}}, \bibinfo {author} {\bibfnamefont {S.}~\bibnamefont {Ramaswamy}},
  \bibinfo {author} {\bibfnamefont {T.~B.}\ \bibnamefont {Liverpool}}, \bibinfo
  {author} {\bibfnamefont {J.}~\bibnamefont {Prost}}, \bibinfo {author}
  {\bibfnamefont {M.}~\bibnamefont {Rao}}, \ and\ \bibinfo {author}
  {\bibfnamefont {R.~A.}\ \bibnamefont {Simha}},\ }\href@noop {} {\bibfield
  {journal} {\bibinfo  {journal} {Reviews of modern physics}\ }\textbf
  {\bibinfo {volume} {85}},\ \bibinfo {pages} {1143} (\bibinfo {year}
  {2013})}\BibitemShut {NoStop}%
\bibitem [{\citenamefont {Bialek}\ \emph {et~al.}(2012)\citenamefont {Bialek},
  \citenamefont {Cavagna}, \citenamefont {Giardina}, \citenamefont {Mora},
  \citenamefont {Silvestri}, \citenamefont {Viale},\ and\ \citenamefont
  {Walczak}}]{bialek2012statistical}%
  \BibitemOpen
  \bibfield  {author} {\bibinfo {author} {\bibfnamefont {W.}~\bibnamefont
  {Bialek}}, \bibinfo {author} {\bibfnamefont {A.}~\bibnamefont {Cavagna}},
  \bibinfo {author} {\bibfnamefont {I.}~\bibnamefont {Giardina}}, \bibinfo
  {author} {\bibfnamefont {T.}~\bibnamefont {Mora}}, \bibinfo {author}
  {\bibfnamefont {E.}~\bibnamefont {Silvestri}}, \bibinfo {author}
  {\bibfnamefont {M.}~\bibnamefont {Viale}}, \ and\ \bibinfo {author}
  {\bibfnamefont {A.~M.}\ \bibnamefont {Walczak}},\ }\href@noop {} {\bibfield
  {journal} {\bibinfo  {journal} {Proceedings of the National Academy of
  Sciences}\ }\textbf {\bibinfo {volume} {109}},\ \bibinfo {pages} {4786}
  (\bibinfo {year} {2012})}\BibitemShut {NoStop}%
\bibitem [{\citenamefont {Toner}\ and\ \citenamefont
  {Tu}(1998)}]{toner1998flocks}%
  \BibitemOpen
  \bibfield  {author} {\bibinfo {author} {\bibfnamefont {J.}~\bibnamefont
  {Toner}}\ and\ \bibinfo {author} {\bibfnamefont {Y.}~\bibnamefont {Tu}},\
  }\href@noop {} {\bibfield  {journal} {\bibinfo  {journal} {Physical review
  E}\ }\textbf {\bibinfo {volume} {58}},\ \bibinfo {pages} {4828} (\bibinfo
  {year} {1998})}\BibitemShut {NoStop}%
\bibitem [{\citenamefont {Elgeti}\ \emph {et~al.}(2015)\citenamefont {Elgeti},
  \citenamefont {Winkler},\ and\ \citenamefont {Gompper}}]{elgeti2015physics}%
  \BibitemOpen
  \bibfield  {author} {\bibinfo {author} {\bibfnamefont {J.}~\bibnamefont
  {Elgeti}}, \bibinfo {author} {\bibfnamefont {R.~G.}\ \bibnamefont {Winkler}},
  \ and\ \bibinfo {author} {\bibfnamefont {G.}~\bibnamefont {Gompper}},\
  }\href@noop {} {\bibfield  {journal} {\bibinfo  {journal} {Reports on
  progress in physics}\ }\textbf {\bibinfo {volume} {78}},\ \bibinfo {pages}
  {056601} (\bibinfo {year} {2015})}\BibitemShut {NoStop}%
\bibitem [{\citenamefont {Yu}\ and\ \citenamefont {Tu}(2022)}]{yu2022energy}%
  \BibitemOpen
  \bibfield  {author} {\bibinfo {author} {\bibfnamefont {Q.}~\bibnamefont
  {Yu}}\ and\ \bibinfo {author} {\bibfnamefont {Y.}~\bibnamefont {Tu}},\
  }\href@noop {} {\bibfield  {journal} {\bibinfo  {journal} {Physical review
  letters}\ }\textbf {\bibinfo {volume} {129}},\ \bibinfo {pages} {278001}
  (\bibinfo {year} {2022})}\BibitemShut {NoStop}%
\bibitem [{\citenamefont {Zhang}\ \emph {et~al.}(2020)\citenamefont {Zhang},
  \citenamefont {Cao}, \citenamefont {Ouyang},\ and\ \citenamefont
  {Tu}}]{sync_energy}%
  \BibitemOpen
  \bibfield  {author} {\bibinfo {author} {\bibfnamefont {D.}~\bibnamefont
  {Zhang}}, \bibinfo {author} {\bibfnamefont {Y.}~\bibnamefont {Cao}}, \bibinfo
  {author} {\bibfnamefont {Q.}~\bibnamefont {Ouyang}}, \ and\ \bibinfo {author}
  {\bibfnamefont {Y.}~\bibnamefont {Tu}},\ }\href@noop {} {\bibfield  {journal}
  {\bibinfo  {journal} {Nature physics}\ }\textbf {\bibinfo {volume} {16}},\
  \bibinfo {pages} {95} (\bibinfo {year} {2020})}\BibitemShut {NoStop}%
\bibitem [{\citenamefont {Acebr{\'o}n}\ \emph {et~al.}(2005)\citenamefont
  {Acebr{\'o}n}, \citenamefont {Bonilla}, \citenamefont {P{\'e}rez~Vicente},
  \citenamefont {Ritort},\ and\ \citenamefont {Spigler}}]{acebron2005kuramoto}%
  \BibitemOpen
  \bibfield  {author} {\bibinfo {author} {\bibfnamefont {J.~A.}\ \bibnamefont
  {Acebr{\'o}n}}, \bibinfo {author} {\bibfnamefont {L.~L.}\ \bibnamefont
  {Bonilla}}, \bibinfo {author} {\bibfnamefont {C.~J.}\ \bibnamefont
  {P{\'e}rez~Vicente}}, \bibinfo {author} {\bibfnamefont {F.}~\bibnamefont
  {Ritort}}, \ and\ \bibinfo {author} {\bibfnamefont {R.}~\bibnamefont
  {Spigler}},\ }\href@noop {} {\bibfield  {journal} {\bibinfo  {journal}
  {Reviews of modern physics}\ }\textbf {\bibinfo {volume} {77}},\ \bibinfo
  {pages} {137} (\bibinfo {year} {2005})}\BibitemShut {NoStop}%
\bibitem [{\citenamefont {Tailleur}\ \emph {et~al.}(2022)\citenamefont
  {Tailleur}, \citenamefont {Gompper}, \citenamefont {Marchetti}, \citenamefont
  {Yeomans},\ and\ \citenamefont {Salomon}}]{tailleur2022active}%
  \BibitemOpen
  \bibfield  {author} {\bibinfo {author} {\bibfnamefont {J.}~\bibnamefont
  {Tailleur}}, \bibinfo {author} {\bibfnamefont {G.}~\bibnamefont {Gompper}},
  \bibinfo {author} {\bibfnamefont {M.~C.}\ \bibnamefont {Marchetti}}, \bibinfo
  {author} {\bibfnamefont {J.~M.}\ \bibnamefont {Yeomans}}, \ and\ \bibinfo
  {author} {\bibfnamefont {C.}~\bibnamefont {Salomon}},\ }\href@noop {} {\emph
  {\bibinfo {title} {Active Matter and Nonequilibrium Statistical Physics:
  Lecture Notes of the Les Houches Summer School: Volume 112, September
  2018}}},\ Vol.\ \bibinfo {volume} {112}\ (\bibinfo  {publisher} {Oxford
  University Press},\ \bibinfo {year} {2022})\BibitemShut {NoStop}%
\bibitem [{\citenamefont {van Zon}\ \emph {et~al.}(2007)\citenamefont {van
  Zon}, \citenamefont {Lubensky}, \citenamefont {Altena},\ and\ \citenamefont
  {ten Wolde}}]{Wolde2007}%
  \BibitemOpen
  \bibfield  {author} {\bibinfo {author} {\bibfnamefont {J.~S.}\ \bibnamefont
  {van Zon}}, \bibinfo {author} {\bibfnamefont {D.~K.}\ \bibnamefont
  {Lubensky}}, \bibinfo {author} {\bibfnamefont {P.~R.}\ \bibnamefont
  {Altena}}, \ and\ \bibinfo {author} {\bibfnamefont {P.~R.}\ \bibnamefont {ten
  Wolde}},\ }\href@noop {} {\bibfield  {journal} {\bibinfo  {journal}
  {Proceedings of the National Academy of Sciences}\ }\textbf {\bibinfo
  {volume} {104}},\ \bibinfo {pages} {7420} (\bibinfo {year}
  {2007})}\BibitemShut {NoStop}%
\bibitem [{\citenamefont {Nagai}\ \emph {et~al.}(2010)\citenamefont {Nagai},
  \citenamefont {Terada},\ and\ \citenamefont {Sasai}}]{Sasai2010}%
  \BibitemOpen
  \bibfield  {author} {\bibinfo {author} {\bibfnamefont {T.}~\bibnamefont
  {Nagai}}, \bibinfo {author} {\bibfnamefont {T.~P.}\ \bibnamefont {Terada}}, \
  and\ \bibinfo {author} {\bibfnamefont {M.}~\bibnamefont {Sasai}},\
  }\href@noop {} {\bibfield  {journal} {\bibinfo  {journal} {Biophysical
  journal}\ }\textbf {\bibinfo {volume} {98}},\ \bibinfo {pages} {2469}
  (\bibinfo {year} {2010})}\BibitemShut {NoStop}%
\bibitem [{\citenamefont {Paijmans}\ \emph {et~al.}(2017)\citenamefont
  {Paijmans}, \citenamefont {Lubensky},\ and\ \citenamefont {ten
  Wolde}}]{Wolde2017}%
  \BibitemOpen
  \bibfield  {author} {\bibinfo {author} {\bibfnamefont {J.}~\bibnamefont
  {Paijmans}}, \bibinfo {author} {\bibfnamefont {D.~K.}\ \bibnamefont
  {Lubensky}}, \ and\ \bibinfo {author} {\bibfnamefont {P.~R.}\ \bibnamefont
  {ten Wolde}},\ }\href@noop {} {\bibfield  {journal} {\bibinfo  {journal}
  {PLoS computational biology}\ }\textbf {\bibinfo {volume} {13}},\ \bibinfo
  {pages} {e1005415} (\bibinfo {year} {2017})}\BibitemShut {NoStop}%
\bibitem [{\citenamefont {Kageyama}\ \emph {et~al.}(2006)\citenamefont
  {Kageyama}, \citenamefont {Nishiwaki}, \citenamefont {Nakajima},
  \citenamefont {Iwasaki}, \citenamefont {Oyama},\ and\ \citenamefont
  {Kondo}}]{Kondo2006}%
  \BibitemOpen
  \bibfield  {author} {\bibinfo {author} {\bibfnamefont {H.}~\bibnamefont
  {Kageyama}}, \bibinfo {author} {\bibfnamefont {T.}~\bibnamefont {Nishiwaki}},
  \bibinfo {author} {\bibfnamefont {M.}~\bibnamefont {Nakajima}}, \bibinfo
  {author} {\bibfnamefont {H.}~\bibnamefont {Iwasaki}}, \bibinfo {author}
  {\bibfnamefont {T.}~\bibnamefont {Oyama}}, \ and\ \bibinfo {author}
  {\bibfnamefont {T.}~\bibnamefont {Kondo}},\ }\href@noop {} {\bibfield
  {journal} {\bibinfo  {journal} {Molecular cell}\ }\textbf {\bibinfo {volume}
  {23}},\ \bibinfo {pages} {161} (\bibinfo {year} {2006})}\BibitemShut
  {NoStop}%
\bibitem [{\citenamefont {Emberly}\ and\ \citenamefont
  {Wingreen}(2006)}]{emberly2006}%
  \BibitemOpen
  \bibfield  {author} {\bibinfo {author} {\bibfnamefont {E.}~\bibnamefont
  {Emberly}}\ and\ \bibinfo {author} {\bibfnamefont {N.~S.}\ \bibnamefont
  {Wingreen}},\ }\href@noop {} {\bibfield  {journal} {\bibinfo  {journal}
  {Physical review letters}\ }\textbf {\bibinfo {volume} {96}},\ \bibinfo
  {pages} {038303} (\bibinfo {year} {2006})}\BibitemShut {NoStop}%
\bibitem [{\citenamefont {Ito}\ \emph {et~al.}(2007)\citenamefont {Ito},
  \citenamefont {Kageyama}, \citenamefont {Mutsuda}, \citenamefont {Nakajima},
  \citenamefont {Oyama},\ and\ \citenamefont {Kondo}}]{Ito2007}%
  \BibitemOpen
  \bibfield  {author} {\bibinfo {author} {\bibfnamefont {H.}~\bibnamefont
  {Ito}}, \bibinfo {author} {\bibfnamefont {H.}~\bibnamefont {Kageyama}},
  \bibinfo {author} {\bibfnamefont {M.}~\bibnamefont {Mutsuda}}, \bibinfo
  {author} {\bibfnamefont {M.}~\bibnamefont {Nakajima}}, \bibinfo {author}
  {\bibfnamefont {T.}~\bibnamefont {Oyama}}, \ and\ \bibinfo {author}
  {\bibfnamefont {T.}~\bibnamefont {Kondo}},\ }\href@noop {} {\bibfield
  {journal} {\bibinfo  {journal} {Nature structural \& molecular biology}\
  }\textbf {\bibinfo {volume} {14}},\ \bibinfo {pages} {1084} (\bibinfo {year}
  {2007})}\BibitemShut {NoStop}%
\bibitem [{\citenamefont {Mori}\ \emph {et~al.}(2007)\citenamefont {Mori},
  \citenamefont {Williams}, \citenamefont {Byrne}, \citenamefont {Qin},
  \citenamefont {Egli}, \citenamefont {Mchaourab}, \citenamefont {Stewart},\
  and\ \citenamefont {Johnson}}]{Johnson2007}%
  \BibitemOpen
  \bibfield  {author} {\bibinfo {author} {\bibfnamefont {T.}~\bibnamefont
  {Mori}}, \bibinfo {author} {\bibfnamefont {D.~R.}\ \bibnamefont {Williams}},
  \bibinfo {author} {\bibfnamefont {M.~O.}\ \bibnamefont {Byrne}}, \bibinfo
  {author} {\bibfnamefont {X.}~\bibnamefont {Qin}}, \bibinfo {author}
  {\bibfnamefont {M.}~\bibnamefont {Egli}}, \bibinfo {author} {\bibfnamefont
  {H.~S.}\ \bibnamefont {Mchaourab}}, \bibinfo {author} {\bibfnamefont {P.~L.}\
  \bibnamefont {Stewart}}, \ and\ \bibinfo {author} {\bibfnamefont {C.~H.}\
  \bibnamefont {Johnson}},\ }\href@noop {} {\bibfield  {journal} {\bibinfo
  {journal} {PLoS biology}\ }\textbf {\bibinfo {volume} {5}},\ \bibinfo {pages}
  {e93} (\bibinfo {year} {2007})}\BibitemShut {NoStop}%
\bibitem [{\citenamefont {Yoda}\ \emph {et~al.}(2007)\citenamefont {Yoda},
  \citenamefont {Eguchi}, \citenamefont {Terada},\ and\ \citenamefont
  {Sasai}}]{Sasai2007}%
  \BibitemOpen
  \bibfield  {author} {\bibinfo {author} {\bibfnamefont {M.}~\bibnamefont
  {Yoda}}, \bibinfo {author} {\bibfnamefont {K.}~\bibnamefont {Eguchi}},
  \bibinfo {author} {\bibfnamefont {T.~P.}\ \bibnamefont {Terada}}, \ and\
  \bibinfo {author} {\bibfnamefont {M.}~\bibnamefont {Sasai}},\ }\href@noop {}
  {\bibfield  {journal} {\bibinfo  {journal} {PloS one}\ }\textbf {\bibinfo
  {volume} {2}},\ \bibinfo {pages} {e408} (\bibinfo {year} {2007})}\BibitemShut
  {NoStop}%
\bibitem [{\citenamefont {Eguchi}\ \emph {et~al.}(2008)\citenamefont {Eguchi},
  \citenamefont {Yoda}, \citenamefont {Terada},\ and\ \citenamefont
  {Sasai}}]{Sasai2008}%
  \BibitemOpen
  \bibfield  {author} {\bibinfo {author} {\bibfnamefont {K.}~\bibnamefont
  {Eguchi}}, \bibinfo {author} {\bibfnamefont {M.}~\bibnamefont {Yoda}},
  \bibinfo {author} {\bibfnamefont {T.~P.}\ \bibnamefont {Terada}}, \ and\
  \bibinfo {author} {\bibfnamefont {M.}~\bibnamefont {Sasai}},\ }\href@noop {}
  {\bibfield  {journal} {\bibinfo  {journal} {Biophysical journal}\ }\textbf
  {\bibinfo {volume} {95}},\ \bibinfo {pages} {1773} (\bibinfo {year}
  {2008})}\BibitemShut {NoStop}%
\bibitem [{\citenamefont {Rust}\ \emph {et~al.}(2007)\citenamefont {Rust},
  \citenamefont {Markson}, \citenamefont {Lane}, \citenamefont {Fisher},\ and\
  \citenamefont {O'shea}}]{rust2007}%
  \BibitemOpen
  \bibfield  {author} {\bibinfo {author} {\bibfnamefont {M.~J.}\ \bibnamefont
  {Rust}}, \bibinfo {author} {\bibfnamefont {J.~S.}\ \bibnamefont {Markson}},
  \bibinfo {author} {\bibfnamefont {W.~S.}\ \bibnamefont {Lane}}, \bibinfo
  {author} {\bibfnamefont {D.~S.}\ \bibnamefont {Fisher}}, \ and\ \bibinfo
  {author} {\bibfnamefont {E.~K.}\ \bibnamefont {O'shea}},\ }\href@noop {}
  {\bibfield  {journal} {\bibinfo  {journal} {Science}\ }\textbf {\bibinfo
  {volume} {318}},\ \bibinfo {pages} {809} (\bibinfo {year}
  {2007})}\BibitemShut {NoStop}%
\bibitem [{\citenamefont {Ma}\ and\ \citenamefont
  {Ranganathan}(2012{\natexlab{a}})}]{ma2012}%
  \BibitemOpen
  \bibfield  {author} {\bibinfo {author} {\bibfnamefont {L.}~\bibnamefont
  {Ma}}\ and\ \bibinfo {author} {\bibfnamefont {R.}~\bibnamefont
  {Ranganathan}},\ }\href@noop {} {\bibfield  {journal} {\bibinfo  {journal}
  {PLoS One}\ }\textbf {\bibinfo {volume} {7}},\ \bibinfo {pages} {e42581}
  (\bibinfo {year} {2012}{\natexlab{a}})}\BibitemShut {NoStop}%
\bibitem [{\citenamefont {Lin}\ \emph {et~al.}(2014)\citenamefont {Lin},
  \citenamefont {Chew}, \citenamefont {Chockanathan},\ and\ \citenamefont
  {Rust}}]{Rust2014}%
  \BibitemOpen
  \bibfield  {author} {\bibinfo {author} {\bibfnamefont {J.}~\bibnamefont
  {Lin}}, \bibinfo {author} {\bibfnamefont {J.}~\bibnamefont {Chew}}, \bibinfo
  {author} {\bibfnamefont {U.}~\bibnamefont {Chockanathan}}, \ and\ \bibinfo
  {author} {\bibfnamefont {M.~J.}\ \bibnamefont {Rust}},\ }\href@noop {}
  {\bibfield  {journal} {\bibinfo  {journal} {Proceedings of the National
  Academy of Sciences}\ }\textbf {\bibinfo {volume} {111}},\ \bibinfo {pages}
  {E3937} (\bibinfo {year} {2014})}\BibitemShut {NoStop}%
\bibitem [{\citenamefont {Behera}\ \emph {et~al.}(2021)\citenamefont {Behera},
  \citenamefont {Junco},\ and\ \citenamefont {Vaikuntanathan}}]{suri2021Kai}%
  \BibitemOpen
  \bibfield  {author} {\bibinfo {author} {\bibfnamefont {A.~K.}\ \bibnamefont
  {Behera}}, \bibinfo {author} {\bibfnamefont {C.~d.}\ \bibnamefont {Junco}}, \
  and\ \bibinfo {author} {\bibfnamefont {S.}~\bibnamefont {Vaikuntanathan}},\
  }\href@noop {} {\bibfield  {journal} {\bibinfo  {journal} {The Journal of
  Physical Chemistry B}\ }\textbf {\bibinfo {volume} {125}},\ \bibinfo {pages}
  {11179} (\bibinfo {year} {2021})}\BibitemShut {NoStop}%
\bibitem [{\citenamefont {Mori}\ \emph {et~al.}(2018)\citenamefont {Mori},
  \citenamefont {Sugiyama}, \citenamefont {Byrne}, \citenamefont {Johnson},
  \citenamefont {Uchihashi},\ and\ \citenamefont {Ando}}]{mori2018}%
  \BibitemOpen
  \bibfield  {author} {\bibinfo {author} {\bibfnamefont {T.}~\bibnamefont
  {Mori}}, \bibinfo {author} {\bibfnamefont {S.}~\bibnamefont {Sugiyama}},
  \bibinfo {author} {\bibfnamefont {M.}~\bibnamefont {Byrne}}, \bibinfo
  {author} {\bibfnamefont {C.~H.}\ \bibnamefont {Johnson}}, \bibinfo {author}
  {\bibfnamefont {T.}~\bibnamefont {Uchihashi}}, \ and\ \bibinfo {author}
  {\bibfnamefont {T.}~\bibnamefont {Ando}},\ }\href@noop {} {\bibfield
  {journal} {\bibinfo  {journal} {Nature Communications}\ }\textbf {\bibinfo
  {volume} {9}},\ \bibinfo {pages} {3245} (\bibinfo {year} {2018})}\BibitemShut
  {NoStop}%
\bibitem [{Note1()}]{Note1}%
  \BibitemOpen
  \bibinfo {note} {We did rescaling $k(\Delta x)^2\to k$ in the continuum
  limit}\BibitemShut {NoStop}%
\bibitem [{\citenamefont {Tom{\'e}}\ and\ \citenamefont
  {de~Oliveira}(2010)}]{tome2010entropy}%
  \BibitemOpen
  \bibfield  {author} {\bibinfo {author} {\bibfnamefont {T.}~\bibnamefont
  {Tom{\'e}}}\ and\ \bibinfo {author} {\bibfnamefont {M.~J.}\ \bibnamefont
  {de~Oliveira}},\ }\href@noop {} {\bibfield  {journal} {\bibinfo  {journal}
  {Physical review E}\ }\textbf {\bibinfo {volume} {82}},\ \bibinfo {pages}
  {021120} (\bibinfo {year} {2010})}\BibitemShut {NoStop}%
\bibitem [{\citenamefont {Zhang}\ and\ \citenamefont
  {Ouyang}(2021)}]{zhang2021entropy}%
  \BibitemOpen
  \bibfield  {author} {\bibinfo {author} {\bibfnamefont {D.}~\bibnamefont
  {Zhang}}\ and\ \bibinfo {author} {\bibfnamefont {Q.}~\bibnamefont {Ouyang}},\
  }\href@noop {} {\bibfield  {journal} {\bibinfo  {journal} {Entropy}\ }\textbf
  {\bibinfo {volume} {23}},\ \bibinfo {pages} {271} (\bibinfo {year}
  {2021})}\BibitemShut {NoStop}%
\bibitem [{Note2()}]{Note2}%
  \BibitemOpen
  \bibinfo {note} {$e_p^*$ and $\alpha ^*$ can be determined from $A_t^*$ and
  the targeted performance $(v,A)$}\BibitemShut {NoStop}%
\bibitem [{\citenamefont {Barato}\ and\ \citenamefont
  {Seifert}(2015)}]{barato2015thermodynamic}%
  \BibitemOpen
  \bibfield  {author} {\bibinfo {author} {\bibfnamefont {A.~C.}\ \bibnamefont
  {Barato}}\ and\ \bibinfo {author} {\bibfnamefont {U.}~\bibnamefont
  {Seifert}},\ }\href@noop {} {\bibfield  {journal} {\bibinfo  {journal}
  {Physical review letters}\ }\textbf {\bibinfo {volume} {114}},\ \bibinfo
  {pages} {158101} (\bibinfo {year} {2015})}\BibitemShut {NoStop}%
\bibitem [{\citenamefont {Horowitz}\ and\ \citenamefont
  {Gingrich}(2020)}]{TUR}%
  \BibitemOpen
  \bibfield  {author} {\bibinfo {author} {\bibfnamefont {J.~M.}\ \bibnamefont
  {Horowitz}}\ and\ \bibinfo {author} {\bibfnamefont {T.~R.}\ \bibnamefont
  {Gingrich}},\ }\href@noop {} {\bibfield  {journal} {\bibinfo  {journal}
  {Nature Physics}\ }\textbf {\bibinfo {volume} {16}},\ \bibinfo {pages} {15}
  (\bibinfo {year} {2020})}\BibitemShut {NoStop}%
\bibitem [{\citenamefont {Lee}\ \emph {et~al.}(2018)\citenamefont {Lee},
  \citenamefont {Hyeon},\ and\ \citenamefont {Jo}}]{lee2018thermodynamic}%
  \BibitemOpen
  \bibfield  {author} {\bibinfo {author} {\bibfnamefont {S.}~\bibnamefont
  {Lee}}, \bibinfo {author} {\bibfnamefont {C.}~\bibnamefont {Hyeon}}, \ and\
  \bibinfo {author} {\bibfnamefont {J.}~\bibnamefont {Jo}},\ }\href@noop {}
  {\bibfield  {journal} {\bibinfo  {journal} {Physical Review E}\ }\textbf
  {\bibinfo {volume} {98}},\ \bibinfo {pages} {032119} (\bibinfo {year}
  {2018})}\BibitemShut {NoStop}%
\bibitem [{\citenamefont {Dieball}\ and\ \citenamefont
  {Godec}(2023)}]{dieball2023direct}%
  \BibitemOpen
  \bibfield  {author} {\bibinfo {author} {\bibfnamefont {C.}~\bibnamefont
  {Dieball}}\ and\ \bibinfo {author} {\bibfnamefont {A.}~\bibnamefont
  {Godec}},\ }\href@noop {} {\bibfield  {journal} {\bibinfo  {journal}
  {Physical Review Letters}\ }\textbf {\bibinfo {volume} {130}},\ \bibinfo
  {pages} {087101} (\bibinfo {year} {2023})}\BibitemShut {NoStop}%
\bibitem [{\citenamefont {Tu}\ and\ \citenamefont
  {Rappel}(2018)}]{Tu2018Adaptation}%
  \BibitemOpen
  \bibfield  {author} {\bibinfo {author} {\bibfnamefont {Y.}~\bibnamefont
  {Tu}}\ and\ \bibinfo {author} {\bibfnamefont {W.-J.}\ \bibnamefont
  {Rappel}},\ }\href {\doibase
  https://doi.org/10.1146/annurev-conmatphys-033117-054046} {\bibfield
  {journal} {\bibinfo  {journal} {Annual Review of Condensed Matter Physics}\
  }\textbf {\bibinfo {volume} {9}},\ \bibinfo {pages} {183} (\bibinfo {year}
  {2018})}\BibitemShut {NoStop}%
\bibitem [{\citenamefont {Lan}\ \emph {et~al.}(2012)\citenamefont {Lan},
  \citenamefont {Sartori}, \citenamefont {Neumann}, \citenamefont {Sourjik},\
  and\ \citenamefont {Tu}}]{lan2012energy}%
  \BibitemOpen
  \bibfield  {author} {\bibinfo {author} {\bibfnamefont {G.}~\bibnamefont
  {Lan}}, \bibinfo {author} {\bibfnamefont {P.}~\bibnamefont {Sartori}},
  \bibinfo {author} {\bibfnamefont {S.}~\bibnamefont {Neumann}}, \bibinfo
  {author} {\bibfnamefont {V.}~\bibnamefont {Sourjik}}, \ and\ \bibinfo
  {author} {\bibfnamefont {Y.}~\bibnamefont {Tu}},\ }\href@noop {} {\bibfield
  {journal} {\bibinfo  {journal} {Nature physics}\ }\textbf {\bibinfo {volume}
  {8}},\ \bibinfo {pages} {422} (\bibinfo {year} {2012})}\BibitemShut {NoStop}%
\bibitem [{\citenamefont {Lan}\ and\ \citenamefont {Tu}(2013)}]{lan2013cost}%
  \BibitemOpen
  \bibfield  {author} {\bibinfo {author} {\bibfnamefont {G.}~\bibnamefont
  {Lan}}\ and\ \bibinfo {author} {\bibfnamefont {Y.}~\bibnamefont {Tu}},\
  }\href@noop {} {\bibfield  {journal} {\bibinfo  {journal} {Journal of The
  Royal Society Interface}\ }\textbf {\bibinfo {volume} {10}},\ \bibinfo
  {pages} {20130489} (\bibinfo {year} {2013})}\BibitemShut {NoStop}%
\bibitem [{\citenamefont {Sartori}\ and\ \citenamefont
  {Tu}(2015)}]{Sartori2015Free}%
  \BibitemOpen
  \bibfield  {author} {\bibinfo {author} {\bibfnamefont {P.}~\bibnamefont
  {Sartori}}\ and\ \bibinfo {author} {\bibfnamefont {Y.}~\bibnamefont {Tu}},\
  }\href {\doibase 10.1103/PhysRevLett.115.118102} {\bibfield  {journal}
  {\bibinfo  {journal} {Phys. Rev. Lett.}\ }\textbf {\bibinfo {volume} {115}},\
  \bibinfo {pages} {118102} (\bibinfo {year} {2015})}\BibitemShut {NoStop}%
\bibitem [{\citenamefont {Cao}\ \emph {et~al.}(2015)\citenamefont {Cao},
  \citenamefont {Wang}, \citenamefont {Ouyang},\ and\ \citenamefont
  {Tu}}]{Cao2015Free-energy}%
  \BibitemOpen
  \bibfield  {author} {\bibinfo {author} {\bibfnamefont {Y.}~\bibnamefont
  {Cao}}, \bibinfo {author} {\bibfnamefont {H.}~\bibnamefont {Wang}}, \bibinfo
  {author} {\bibfnamefont {Q.}~\bibnamefont {Ouyang}}, \ and\ \bibinfo {author}
  {\bibfnamefont {Y.}~\bibnamefont {Tu}},\ }\href {\doibase 10.1038/nphys3412}
  {\bibfield  {journal} {\bibinfo  {journal} {Nature Physics}\ }\textbf
  {\bibinfo {volume} {11}},\ \bibinfo {pages} {772} (\bibinfo {year}
  {2015})}\BibitemShut {NoStop}%
\bibitem [{\citenamefont {Fei}\ \emph {et~al.}(2018)\citenamefont {Fei},
  \citenamefont {Cao}, \citenamefont {Ouyang},\ and\ \citenamefont
  {Tu}}]{Fei2018Design}%
  \BibitemOpen
  \bibfield  {author} {\bibinfo {author} {\bibfnamefont {C.}~\bibnamefont
  {Fei}}, \bibinfo {author} {\bibfnamefont {Y.}~\bibnamefont {Cao}}, \bibinfo
  {author} {\bibfnamefont {Q.}~\bibnamefont {Ouyang}}, \ and\ \bibinfo {author}
  {\bibfnamefont {Y.}~\bibnamefont {Tu}},\ }\href {\doibase
  10.1038/s41467-018-03826-4} {\bibfield  {journal} {\bibinfo  {journal}
  {Nature Communications}\ }\textbf {\bibinfo {volume} {9}},\ \bibinfo {pages}
  {1434} (\bibinfo {year} {2018})}\BibitemShut {NoStop}%
\bibitem [{\citenamefont {Hathcock}\ \emph {et~al.}(2023)\citenamefont
  {Hathcock}, \citenamefont {Yu}, \citenamefont {Mello}, \citenamefont {Amin},
  \citenamefont {Hazelbauer},\ and\ \citenamefont
  {Tu}}]{Hathcock2023Nonequilibrium}%
  \BibitemOpen
  \bibfield  {author} {\bibinfo {author} {\bibfnamefont {D.}~\bibnamefont
  {Hathcock}}, \bibinfo {author} {\bibfnamefont {Q.}~\bibnamefont {Yu}},
  \bibinfo {author} {\bibfnamefont {B.~A.}\ \bibnamefont {Mello}}, \bibinfo
  {author} {\bibfnamefont {D.~N.}\ \bibnamefont {Amin}}, \bibinfo {author}
  {\bibfnamefont {G.~L.}\ \bibnamefont {Hazelbauer}}, \ and\ \bibinfo {author}
  {\bibfnamefont {Y.}~\bibnamefont {Tu}},\ }\href {\doibase
  10.1073/pnas.2303115120} {\bibfield  {journal} {\bibinfo  {journal}
  {Proceedings of the National Academy of Sciences}\ }\textbf {\bibinfo
  {volume} {120}},\ \bibinfo {pages} {e2303115120} (\bibinfo {year} {2023})},\
  \Eprint
  {http://arxiv.org/abs/https://www.pnas.org/doi/pdf/10.1073/pnas.2303115120}
  {https://www.pnas.org/doi/pdf/10.1073/pnas.2303115120} \BibitemShut {NoStop}%
\bibitem [{\citenamefont {Hathcock}\ \emph {et~al.}(2024)\citenamefont
  {Hathcock}, \citenamefont {Yu},\ and\ \citenamefont
  {Tu}}]{Hathcock2024Time-reversal}%
  \BibitemOpen
  \bibfield  {author} {\bibinfo {author} {\bibfnamefont {D.}~\bibnamefont
  {Hathcock}}, \bibinfo {author} {\bibfnamefont {Q.}~\bibnamefont {Yu}}, \ and\
  \bibinfo {author} {\bibfnamefont {Y.}~\bibnamefont {Tu}},\ }\href {\doibase
  10.1038/s41467-024-52799-0} {\bibfield  {journal} {\bibinfo  {journal}
  {Nature Communications}\ }\textbf {\bibinfo {volume} {15}},\ \bibinfo {pages}
  {8892} (\bibinfo {year} {2024})}\BibitemShut {NoStop}%
\bibitem [{\citenamefont {Ma}\ and\ \citenamefont
  {Ranganathan}(2012{\natexlab{b}})}]{KaiA_affinity}%
  \BibitemOpen
  \bibfield  {author} {\bibinfo {author} {\bibfnamefont {L.}~\bibnamefont
  {Ma}}\ and\ \bibinfo {author} {\bibfnamefont {R.}~\bibnamefont
  {Ranganathan}},\ }\href@noop {} {\bibfield  {journal} {\bibinfo  {journal}
  {PLOS One}\ } (\bibinfo {year} {2012}{\natexlab{b}})}\BibitemShut {NoStop}%
\bibitem [{\citenamefont {Takigawa-Imamura}\ and\ \citenamefont
  {Mochizuki}(2006)}]{takigawa2006predicting}%
  \BibitemOpen
  \bibfield  {author} {\bibinfo {author} {\bibfnamefont {H.}~\bibnamefont
  {Takigawa-Imamura}}\ and\ \bibinfo {author} {\bibfnamefont {A.}~\bibnamefont
  {Mochizuki}},\ }\href@noop {} {\bibfield  {journal} {\bibinfo  {journal}
  {Journal of biological rhythms}\ }\textbf {\bibinfo {volume} {21}},\ \bibinfo
  {pages} {405} (\bibinfo {year} {2006})}\BibitemShut {NoStop}%
\bibitem [{\citenamefont {Clodong}\ \emph {et~al.}(2007)\citenamefont
  {Clodong}, \citenamefont {D{\"u}hring}, \citenamefont {Kronk}, \citenamefont
  {Wilde}, \citenamefont {Axmann}, \citenamefont {Herzel},\ and\ \citenamefont
  {Kollmann}}]{clodong2007functioning}%
  \BibitemOpen
  \bibfield  {author} {\bibinfo {author} {\bibfnamefont {S.}~\bibnamefont
  {Clodong}}, \bibinfo {author} {\bibfnamefont {U.}~\bibnamefont
  {D{\"u}hring}}, \bibinfo {author} {\bibfnamefont {L.}~\bibnamefont {Kronk}},
  \bibinfo {author} {\bibfnamefont {A.}~\bibnamefont {Wilde}}, \bibinfo
  {author} {\bibfnamefont {I.}~\bibnamefont {Axmann}}, \bibinfo {author}
  {\bibfnamefont {H.}~\bibnamefont {Herzel}}, \ and\ \bibinfo {author}
  {\bibfnamefont {M.}~\bibnamefont {Kollmann}},\ }\href@noop {} {\bibfield
  {journal} {\bibinfo  {journal} {Molecular systems biology}\ }\textbf
  {\bibinfo {volume} {3}},\ \bibinfo {pages} {90} (\bibinfo {year}
  {2007})}\BibitemShut {NoStop}%
\bibitem [{\citenamefont {Miyoshi}\ \emph {et~al.}(2007)\citenamefont
  {Miyoshi}, \citenamefont {Nakayama}, \citenamefont {Kaizu}, \citenamefont
  {Iwasaki},\ and\ \citenamefont {Tomita}}]{miyoshi2007}%
  \BibitemOpen
  \bibfield  {author} {\bibinfo {author} {\bibfnamefont {F.}~\bibnamefont
  {Miyoshi}}, \bibinfo {author} {\bibfnamefont {Y.}~\bibnamefont {Nakayama}},
  \bibinfo {author} {\bibfnamefont {K.}~\bibnamefont {Kaizu}}, \bibinfo
  {author} {\bibfnamefont {H.}~\bibnamefont {Iwasaki}}, \ and\ \bibinfo
  {author} {\bibfnamefont {M.}~\bibnamefont {Tomita}},\ }\href@noop {}
  {\bibfield  {journal} {\bibinfo  {journal} {Journal of Biological Rhythms}\
  }\textbf {\bibinfo {volume} {22}},\ \bibinfo {pages} {69} (\bibinfo {year}
  {2007})}\BibitemShut {NoStop}%
\bibitem [{\citenamefont {Axmann}\ \emph {et~al.}(2007)\citenamefont {Axmann},
  \citenamefont {Legewie},\ and\ \citenamefont {Herzel}}]{axmann2007minimal}%
  \BibitemOpen
  \bibfield  {author} {\bibinfo {author} {\bibfnamefont {I.~M.}\ \bibnamefont
  {Axmann}}, \bibinfo {author} {\bibfnamefont {S.}~\bibnamefont {Legewie}}, \
  and\ \bibinfo {author} {\bibfnamefont {H.}~\bibnamefont {Herzel}},\
  }\href@noop {} {\bibfield  {journal} {\bibinfo  {journal} {Genome
  Informatics}\ }\textbf {\bibinfo {volume} {18}},\ \bibinfo {pages} {54}
  (\bibinfo {year} {2007})}\BibitemShut {NoStop}%
\bibitem [{\citenamefont {Nakajima}\ \emph {et~al.}(2010)\citenamefont
  {Nakajima}, \citenamefont {Ito},\ and\ \citenamefont
  {Kondo}}]{nakajima2010vitro}%
  \BibitemOpen
  \bibfield  {author} {\bibinfo {author} {\bibfnamefont {M.}~\bibnamefont
  {Nakajima}}, \bibinfo {author} {\bibfnamefont {H.}~\bibnamefont {Ito}}, \
  and\ \bibinfo {author} {\bibfnamefont {T.}~\bibnamefont {Kondo}},\
  }\href@noop {} {\bibfield  {journal} {\bibinfo  {journal} {FEBS letters}\
  }\textbf {\bibinfo {volume} {584}},\ \bibinfo {pages} {898} (\bibinfo {year}
  {2010})}\BibitemShut {NoStop}%
\bibitem [{\citenamefont {Chavan}\ \emph {et~al.}(2021)\citenamefont {Chavan},
  \citenamefont {Swan}, \citenamefont {Heisler}, \citenamefont {Sancar},
  \citenamefont {Ernst}, \citenamefont {Fang}, \citenamefont {Palacios},
  \citenamefont {Spangler}, \citenamefont {Bagshaw}, \citenamefont {Tripathi}
  \emph {et~al.}}]{chavan2021reconstitution}%
  \BibitemOpen
  \bibfield  {author} {\bibinfo {author} {\bibfnamefont {A.~G.}\ \bibnamefont
  {Chavan}}, \bibinfo {author} {\bibfnamefont {J.~A.}\ \bibnamefont {Swan}},
  \bibinfo {author} {\bibfnamefont {J.}~\bibnamefont {Heisler}}, \bibinfo
  {author} {\bibfnamefont {C.}~\bibnamefont {Sancar}}, \bibinfo {author}
  {\bibfnamefont {D.~C.}\ \bibnamefont {Ernst}}, \bibinfo {author}
  {\bibfnamefont {M.}~\bibnamefont {Fang}}, \bibinfo {author} {\bibfnamefont
  {J.~G.}\ \bibnamefont {Palacios}}, \bibinfo {author} {\bibfnamefont {R.~K.}\
  \bibnamefont {Spangler}}, \bibinfo {author} {\bibfnamefont {C.~R.}\
  \bibnamefont {Bagshaw}}, \bibinfo {author} {\bibfnamefont {S.}~\bibnamefont
  {Tripathi}},  \emph {et~al.},\ }\href@noop {} {\bibfield  {journal} {\bibinfo
   {journal} {Science}\ }\textbf {\bibinfo {volume} {374}},\ \bibinfo {pages}
  {eabd4453} (\bibinfo {year} {2021})}\BibitemShut {NoStop}%
\bibitem [{\citenamefont {Bonsall}\ and\ \citenamefont
  {Wright}(2012)}]{Bonsall2012Altruism}%
  \BibitemOpen
  \bibfield  {author} {\bibinfo {author} {\bibfnamefont {M.~B.}\ \bibnamefont
  {Bonsall}}\ and\ \bibinfo {author} {\bibfnamefont {A.~E.}\ \bibnamefont
  {Wright}},\ }\href {\doibase https://doi.org/10.1002/ece3.206} {\bibfield
  {journal} {\bibinfo  {journal} {Ecology and Evolution}\ }\textbf {\bibinfo
  {volume} {2}},\ \bibinfo {pages} {515} (\bibinfo {year} {2012})},\ \Eprint
  {http://arxiv.org/abs/https://onlinelibrary.wiley.com/doi/pdf/10.1002/ece3.206}
  {https://onlinelibrary.wiley.com/doi/pdf/10.1002/ece3.206} \BibitemShut
  {NoStop}%
\bibitem [{\citenamefont {Kreider}\ \emph {et~al.}(2022)\citenamefont
  {Kreider}, \citenamefont {Janzen}, \citenamefont {Bernadou}, \citenamefont
  {Elsner}, \citenamefont {Kramer},\ and\ \citenamefont
  {Weissing}}]{Kreider2022resource}%
  \BibitemOpen
  \bibfield  {author} {\bibinfo {author} {\bibfnamefont {J.~J.}\ \bibnamefont
  {Kreider}}, \bibinfo {author} {\bibfnamefont {T.}~\bibnamefont {Janzen}},
  \bibinfo {author} {\bibfnamefont {A.}~\bibnamefont {Bernadou}}, \bibinfo
  {author} {\bibfnamefont {D.}~\bibnamefont {Elsner}}, \bibinfo {author}
  {\bibfnamefont {B.~H.}\ \bibnamefont {Kramer}}, \ and\ \bibinfo {author}
  {\bibfnamefont {F.~J.}\ \bibnamefont {Weissing}},\ }\href {\doibase
  10.1038/s41467-022-35038-2} {\bibfield  {journal} {\bibinfo  {journal}
  {Nature Communications}\ }\textbf {\bibinfo {volume} {13}},\ \bibinfo {pages}
  {7232} (\bibinfo {year} {2022})}\BibitemShut {NoStop}%
\bibitem [{\citenamefont {Mott}\ and\ \citenamefont
  {Berger}(2007)}]{mott2007dna}%
  \BibitemOpen
  \bibfield  {author} {\bibinfo {author} {\bibfnamefont {M.~L.}\ \bibnamefont
  {Mott}}\ and\ \bibinfo {author} {\bibfnamefont {J.~M.}\ \bibnamefont
  {Berger}},\ }\href@noop {} {\bibfield  {journal} {\bibinfo  {journal} {Nature
  Reviews Microbiology}\ }\textbf {\bibinfo {volume} {5}},\ \bibinfo {pages}
  {343} (\bibinfo {year} {2007})}\BibitemShut {NoStop}%
\bibitem [{\citenamefont {Fu}\ \emph {et~al.}(2023)\citenamefont {Fu},
  \citenamefont {Xiao},\ and\ \citenamefont {Jun}}]{fu2023bacterial}%
  \BibitemOpen
  \bibfield  {author} {\bibinfo {author} {\bibfnamefont {H.}~\bibnamefont
  {Fu}}, \bibinfo {author} {\bibfnamefont {F.}~\bibnamefont {Xiao}}, \ and\
  \bibinfo {author} {\bibfnamefont {S.}~\bibnamefont {Jun}},\ }\href@noop {}
  {\bibfield  {journal} {\bibinfo  {journal} {PRX life}\ }\textbf {\bibinfo
  {volume} {1}},\ \bibinfo {pages} {013011} (\bibinfo {year}
  {2023})}\BibitemShut {NoStop}%
\bibitem [{\citenamefont {Antebi}\ \emph {et~al.}(2017)\citenamefont {Antebi},
  \citenamefont {Linton}, \citenamefont {Klumpe}, \citenamefont {Bintu},
  \citenamefont {Gong}, \citenamefont {Su}, \citenamefont {McCardell},\ and\
  \citenamefont {Elowitz}}]{Antebi2017combinatorial}%
  \BibitemOpen
  \bibfield  {author} {\bibinfo {author} {\bibfnamefont {Y.~E.}\ \bibnamefont
  {Antebi}}, \bibinfo {author} {\bibfnamefont {J.~M.}\ \bibnamefont {Linton}},
  \bibinfo {author} {\bibfnamefont {H.}~\bibnamefont {Klumpe}}, \bibinfo
  {author} {\bibfnamefont {B.}~\bibnamefont {Bintu}}, \bibinfo {author}
  {\bibfnamefont {M.}~\bibnamefont {Gong}}, \bibinfo {author} {\bibfnamefont
  {C.}~\bibnamefont {Su}}, \bibinfo {author} {\bibfnamefont {R.}~\bibnamefont
  {McCardell}}, \ and\ \bibinfo {author} {\bibfnamefont {M.~B.}\ \bibnamefont
  {Elowitz}},\ }\href {https://doi.org/10.1016/j.cell.2017.08.015} {\bibfield
  {journal} {\bibinfo  {journal} {Cell}\ }\textbf {\bibinfo {volume} {170}},\
  \bibinfo {pages} {1184} (\bibinfo {year} {2017})}\BibitemShut {NoStop}%
\bibitem [{\citenamefont {Su}\ \emph {et~al.}(2022)\citenamefont {Su},
  \citenamefont {Murugan}, \citenamefont {Linton}, \citenamefont {Yeluri},
  \citenamefont {Bois}, \citenamefont {Klumpe}, \citenamefont {Langley},
  \citenamefont {Antebi},\ and\ \citenamefont {Elowitz}}]{su2022ligand}%
  \BibitemOpen
  \bibfield  {author} {\bibinfo {author} {\bibfnamefont {C.~J.}\ \bibnamefont
  {Su}}, \bibinfo {author} {\bibfnamefont {A.}~\bibnamefont {Murugan}},
  \bibinfo {author} {\bibfnamefont {J.~M.}\ \bibnamefont {Linton}}, \bibinfo
  {author} {\bibfnamefont {A.}~\bibnamefont {Yeluri}}, \bibinfo {author}
  {\bibfnamefont {J.}~\bibnamefont {Bois}}, \bibinfo {author} {\bibfnamefont
  {H.}~\bibnamefont {Klumpe}}, \bibinfo {author} {\bibfnamefont {M.~A.}\
  \bibnamefont {Langley}}, \bibinfo {author} {\bibfnamefont {Y.~E.}\
  \bibnamefont {Antebi}}, \ and\ \bibinfo {author} {\bibfnamefont {M.~B.}\
  \bibnamefont {Elowitz}},\ }\href@noop {} {\bibfield  {journal} {\bibinfo
  {journal} {Cell systems}\ }\textbf {\bibinfo {volume} {13}},\ \bibinfo
  {pages} {408} (\bibinfo {year} {2022})}\BibitemShut {NoStop}%
\bibitem [{\citenamefont {Klumpe}\ \emph {et~al.}(2022)\citenamefont {Klumpe},
  \citenamefont {Langley}, \citenamefont {Linton}, \citenamefont {Su},
  \citenamefont {Antebi},\ and\ \citenamefont {Elowitz}}]{klumpe2022context}%
  \BibitemOpen
  \bibfield  {author} {\bibinfo {author} {\bibfnamefont {H.~E.}\ \bibnamefont
  {Klumpe}}, \bibinfo {author} {\bibfnamefont {M.~A.}\ \bibnamefont {Langley}},
  \bibinfo {author} {\bibfnamefont {J.~M.}\ \bibnamefont {Linton}}, \bibinfo
  {author} {\bibfnamefont {C.~J.}\ \bibnamefont {Su}}, \bibinfo {author}
  {\bibfnamefont {Y.~E.}\ \bibnamefont {Antebi}}, \ and\ \bibinfo {author}
  {\bibfnamefont {M.~B.}\ \bibnamefont {Elowitz}},\ }\href@noop {} {\bibfield
  {journal} {\bibinfo  {journal} {Cell systems}\ }\textbf {\bibinfo {volume}
  {13}},\ \bibinfo {pages} {388} (\bibinfo {year} {2022})}\BibitemShut
  {NoStop}%
\bibitem [{\citenamefont {Klumpe}\ \emph {et~al.}(2023)\citenamefont {Klumpe},
  \citenamefont {Garcia-Ojalvo}, \citenamefont {Elowitz},\ and\ \citenamefont
  {Antebi}}]{klumpe2023computational}%
  \BibitemOpen
  \bibfield  {author} {\bibinfo {author} {\bibfnamefont {H.~E.}\ \bibnamefont
  {Klumpe}}, \bibinfo {author} {\bibfnamefont {J.}~\bibnamefont
  {Garcia-Ojalvo}}, \bibinfo {author} {\bibfnamefont {M.~B.}\ \bibnamefont
  {Elowitz}}, \ and\ \bibinfo {author} {\bibfnamefont {Y.~E.}\ \bibnamefont
  {Antebi}},\ }\href@noop {} {\bibfield  {journal} {\bibinfo  {journal} {Cell
  Systems}\ }\textbf {\bibinfo {volume} {14}},\ \bibinfo {pages} {430}
  (\bibinfo {year} {2023})}\BibitemShut {NoStop}%
\end{thebibliography}%


\begin{thebibliography}{4}%
\makeatletter
\providecommand \@ifxundefined [1]{%
 \@ifx{#1\undefined}
}%
\providecommand \@ifnum [1]{%
 \ifnum #1\expandafter \@firstoftwo
 \else \expandafter \@secondoftwo
 \fi
}%
\providecommand \@ifx [1]{%
 \ifx #1\expandafter \@firstoftwo
 \else \expandafter \@secondoftwo
 \fi
}%
\providecommand \natexlab [1]{#1}%
\providecommand \enquote  [1]{``#1''}%
\providecommand \bibnamefont  [1]{#1}%
\providecommand \bibfnamefont [1]{#1}%
\providecommand \citenamefont [1]{#1}%
\providecommand \href@noop [0]{\@secondoftwo}%
\providecommand \href [0]{\begingroup \@sanitize@url \@href}%
\providecommand \@href[1]{\@@startlink{#1}\@@href}%
\providecommand \@@href[1]{\endgroup#1\@@endlink}%
\providecommand \@sanitize@url [0]{\catcode `\\12\catcode `\$12\catcode
  `\&12\catcode `\#12\catcode `\^12\catcode `\_12\catcode `\%12\relax}%
\providecommand \@@startlink[1]{}%
\providecommand \@@endlink[0]{}%
\providecommand \url  [0]{\begingroup\@sanitize@url \@url }%
\providecommand \@url [1]{\endgroup\@href {#1}{\urlprefix }}%
\providecommand \urlprefix  [0]{URL }%
\providecommand \Eprint [0]{\href }%
\providecommand \doibase [0]{http://dx.doi.org/}%
\providecommand \selectlanguage [0]{\@gobble}%
\providecommand \bibinfo  [0]{\@secondoftwo}%
\providecommand \bibfield  [0]{\@secondoftwo}%
\providecommand \translation [1]{[#1]}%
\providecommand \BibitemOpen [0]{}%
\providecommand \bibitemStop [0]{}%
\providecommand \bibitemNoStop [0]{.\EOS\space}%
\providecommand \EOS [0]{\spacefactor3000\relax}%
\providecommand \BibitemShut  [1]{\csname bibitem#1\endcsname}%
\let\auto@bib@innerbib\@empty
\bibitem [{\citenamefont {Reif}(1965)}]{statistical_mechanics}%
  \BibitemOpen
  \bibfield  {author} {\bibinfo {author} {\bibfnamefont {F.}~\bibnamefont
  {Reif}},\ }\href@noop {} {\emph {\bibinfo {title} {Fundamentals of
  Statistical and Thermal Physics}}}\ (\bibinfo  {publisher} {McGraw–Hill},\
  \bibinfo {year} {1965})\BibitemShut {NoStop}%
\bibitem [{\citenamefont {Zhang}\ \emph {et~al.}(2020)\citenamefont {Zhang},
  \citenamefont {Cao}, \citenamefont {Ouyang},\ and\ \citenamefont
  {Tu}}]{sync_energy}%
  \BibitemOpen
  \bibfield  {author} {\bibinfo {author} {\bibfnamefont {D.}~\bibnamefont
  {Zhang}}, \bibinfo {author} {\bibfnamefont {Y.}~\bibnamefont {Cao}}, \bibinfo
  {author} {\bibfnamefont {Q.}~\bibnamefont {Ouyang}}, \ and\ \bibinfo {author}
  {\bibfnamefont {Y.}~\bibnamefont {Tu}},\ }\href@noop {} {\bibfield  {journal}
  {\bibinfo  {journal} {Nature physics}\ }\textbf {\bibinfo {volume} {16}},\
  \bibinfo {pages} {95} (\bibinfo {year} {2020})}\BibitemShut {NoStop}%
\bibitem [{\citenamefont {Horowitz}\ and\ \citenamefont
  {Gingrich}(2020)}]{TUR}%
  \BibitemOpen
  \bibfield  {author} {\bibinfo {author} {\bibfnamefont {J.~M.}\ \bibnamefont
  {Horowitz}}\ and\ \bibinfo {author} {\bibfnamefont {T.~R.}\ \bibnamefont
  {Gingrich}},\ }\href@noop {} {\bibfield  {journal} {\bibinfo  {journal}
  {Nature Physics}\ }\textbf {\bibinfo {volume} {16}},\ \bibinfo {pages} {15}
  (\bibinfo {year} {2020})}\BibitemShut {NoStop}%
\bibitem [{\citenamefont {Lee}\ \emph {et~al.}(2018)\citenamefont {Lee},
  \citenamefont {Hyeon},\ and\ \citenamefont {Jo}}]{lee2018thermodynamic}%
  \BibitemOpen
  \bibfield  {author} {\bibinfo {author} {\bibfnamefont {S.}~\bibnamefont
  {Lee}}, \bibinfo {author} {\bibfnamefont {C.}~\bibnamefont {Hyeon}}, \ and\
  \bibinfo {author} {\bibfnamefont {J.}~\bibnamefont {Jo}},\ }\href@noop {}
  {\bibfield  {journal} {\bibinfo  {journal} {Physical Review E}\ }\textbf
  {\bibinfo {volume} {98}},\ \bibinfo {pages} {032119} (\bibinfo {year}
  {2018})}\BibitemShut {NoStop}%
\end{thebibliography}%
\end{document}


\title{Supplementary information for ``An altruistic resource-sharing mechanism for synchronization:\\ The energy-speed-accuracy tradeoff"}
\maketitle
\section{The simplest case with $N=2,M=1$}

The governing Fokker-Planck equation for $P(x_1,x_2)$ is:
\begin{align}
\frac{\partial P( x_1, x_2,t)}{\partial t}=
-\frac{\partial}{\partial x_1}\left(k_pe_pp_1( x_1, x_2)P-\frac{\partial}{\partial x_1}k_pp_1P
\right) \notag
-\frac{\partial}{\partial x_2}\left(k_pe_pp_2( x_1, x_2)P-\frac{\partial}{\partial x_2}k_pp_2P
\right), 
\end{align}
with
\begin{equation}
    p_1=\frac{e^{-\alpha( x_1- x_2)/2}}{e^{-\alpha( x_1- x_2)/2}+e^{\alpha( x_1- x_2)/2}},
    p_2=\frac{e^{\alpha( x_1- x_2)/2}}{e^{-\alpha( x_1- x_2)/2}+e^{\alpha( x_1- x_2)/2}},
\end{equation}

Rearrange the variables as:
\begin{equation}
\left\{
\begin{aligned}
 u&= x_1- x_2, \\
\bar{x}&=\frac{ x_1+ x_2}{2},
\end{aligned}\right.
\end{equation}
or
\begin{equation}
\begin{pmatrix}
     u \\ \bar{x}
\end{pmatrix}=
\begin{pmatrix}
    1 & -1\\
    \frac12 & \frac12\\
\end{pmatrix}
\begin{pmatrix}
     x_1 \\  x_2
\end{pmatrix}
\equiv \mathbf{M} \begin{pmatrix}
     x_1 \\  x_2
\end{pmatrix}.
\end{equation}
The derivatives satisfy
\begin{equation}
\begin{pmatrix}
    \frac{\partial}{\partial x_1} \\ \frac{\partial}{\partial x_2}
\end{pmatrix}
=
\mathbf{M^T}
\begin{pmatrix}
    \frac{\partial}{\partial u} \\ \frac{\partial}{\partial\bar{x}}
\end{pmatrix}
.
\end{equation}

First, we can calculate the mean velocity of the averaged position: $v\equiv \d\langle \bar{x}\rangle/\d t$ by
\begin{equation}
\frac{\d\langle \bar{x}\rangle}{\d}
=\frac12\int (x_1+x_2)\frac{\partial P}{\partial t}\d x_1\d x_2
=\frac12ke_p\int(p_1+p_2)P\d x_1\d x_2
=\frac12ke_p.
\end{equation}
This result reveals that $v$ is a constant independent of $\alpha$, as expected.

The steady state distribution $P_s(x_1,x_2)$ can be solved in the new coordinates $ u$ and $\bar{x}$. The Fokker-Planck equation is equivalent to 

\begin{align}
\frac{\partial P(u, \bar{x},t)}{\partial t}=
&-\frac{\partial}{\partial u}\left(
-k_pe_p\tanh{(\frac{\alpha u}2)}P
\right) 
-\frac{\partial}{\partial\bar{x}}\left(
\frac12k_p(e_p-\frac{\alpha}{(e^{-\frac{\alpha u}{2}}+e^{\frac{\alpha u}{2}})^2})P
\right)\notag\\
&+k_p\begin{pmatrix}
    \frac{\partial}{\partial x_1} , \frac{\partial}{\partial x_2}
\end{pmatrix}
\begin{pmatrix}
    p_1 & \\
     & p_2\\
\end{pmatrix}
\begin{pmatrix}
    \frac{\partial}{\partial x_1} \\ \frac{\partial}{\partial x_2}
\end{pmatrix}P.
\end{align}
Since
\begin{equation}
\begin{pmatrix}
    \frac{\partial}{\partial x_1} , \frac{\partial}{\partial x_2}
\end{pmatrix}
\begin{pmatrix}
    p_1 & \\
     & p_2\\
\end{pmatrix}
\begin{pmatrix}
    \frac{\partial}{\partial x_1} \\ \frac{\partial}{\partial x_2}
\end{pmatrix}
=
\begin{pmatrix}
    \frac{\partial}{\partial u} , \frac{\partial}{\partial\bar{x}}
\end{pmatrix}
\mathbf{M}
\begin{pmatrix}
    p_1 & \\
     & p_2\\
\end{pmatrix}
\mathbf{M^T}
\begin{pmatrix}
    \frac{\partial}{\partial u} \\ \frac{\partial}{\partial\bar{x}}
\end{pmatrix},
\end{equation}
and
\begin{equation}
\mathbf{M}
\begin{pmatrix}
    p_1 & \\
     & p_2\\
\end{pmatrix}
\mathbf{M^T}
=
\begin{pmatrix}
    1 & \frac12\tanh{\frac{\alpha u}{2}}\\
    \frac12\tanh{\frac{\alpha u}{2}} & \frac14\\
\end{pmatrix},
\end{equation}
the second-order terms are
\begin{equation}
\begin{pmatrix}
    \frac{\partial}{\partial x_1} , \frac{\partial}{\partial x_2}
\end{pmatrix}
\begin{pmatrix}
    p_1 & \\
     & p_2\\
\end{pmatrix}
\begin{pmatrix}
    \frac{\partial}{\partial x_1} \\ \frac{\partial}{\partial x_2}
\end{pmatrix}P
=[\frac{\partial^2}{\partial u^2}+\frac14\frac{\partial^2}{\partial\bar{x}^2}+
\frac12\frac{\partial}{\partial\bar{x}}(
\tanh{\frac{\alpha u}{2}}
\frac{\partial}{\partial u}
+
\frac{\partial}{\partial u}
\tanh{\frac{\alpha u}{2}}
)
]P.
\end{equation}
We aim to find the steady-state solution by using an ansatz: $P_s(x_1,x_2)=P_s(u)$, and we obtain the equation for $P_s(u)$:
\begin{equation}
-k_pe_p\tanh{\frac{\alpha u}2}\times P-k_p\frac{\partial P}{\partial  u}=0, P( u)\propto\exp{(-\frac{2e_p}{\alpha}\ln\cosh{\frac{\alpha u}2})},
\end{equation}
from which we can define an effective energy landscape $E_{eff}( u)$:
\begin{equation}
E_{eff}( u)\equiv-\ln{P( u)}=2\frac{e_p}{\alpha}\ln\cosh{\frac{\alpha u}2}.
\end{equation}

\section{The continuum model in thermodynamic limit}
In this section we derive the analytical expression of $\rho_s(u)$ in the thermodynamic limit.
By denoting the number of activators that bind to $i$-th agent as $m_i$, 
the probability of how the activators are arranged on the agents satisfies detailed balance and follows a canonical distribution \cite{statistical_mechanics}, i.e., the probability of a specific arrangement
$\{m_1,m_2,\ldots,m_N\}$
is proportional to 
$(M!/\prod_jm_j!)\e^{-E_{tot}}$ 
with the prefactor describing the degeneracy (ways of the arrangement) and the total affinity potential $E_{tot}=\sum_jm_jE(x_j)$,
so that $p_i$ is determined formally by
\begin{align} 
p_i=\sum_{\{m_1,m_2,\ldots,m_N\}}m_i\cdot
\frac{M!}Z\prod_j\frac1{m_j!}\e^{-m_jE_j},
\label{eq:pi}
\end{align}
where 
$Z=\sum_{\{m_1,m_2,\ldots,m_N\}}M!\prod_j(m_j!)^{-1}\e^{-m_jE(x_j)}$ is the partition function and the summation runs over all possible arrangements 
$\{m_1,m_2,\ldots,m_N\}$ with the constraint $0\leq m_i \leq g$ and $\sum_jm_j=M$.
The partition function $Z$ is to ensure normalization, i.e., $\sum_i^N p_i=M$.

The analytical expression of $Z$ can be calculated as follows.
By introducing a notation of ``partial partition function"
\begin{equation}
Z_i(M-m_i) \equiv{\sum_{\{m_1,m_2,\ldots,m_N\}}} (g-m_i)!
\prod_{j\neq i}\frac1{m_j!}(\e^{-E_j})^{m_j},
\end{equation}
where the product runs except $i$-th agent and $\sum_{j\neq i}m_j=M-m_i$.
In this way, the partition function can be written as
\begin{equation}
    Z=\sum_{m_i=0}^g \frac{M!}{g!}Z_i(M-m_i) C_g^{m_i}\e^{-m_iE_i}.
\end{equation}
with $C_g^{m_i}=g!/(m_i!(g-m_i)!)$. 
Given $m_i<g\ll M$, we can expand $Z_i(M-m_i)$ by
\[
    \ln Z_i(M-m_i)=\ln Z_i(M)-m_i\frac{\partial \ln Z_i }{\partial M}, \text{or  } 
    Z_i(M-m_i)=Z_i(M)\e^{-m_i\zeta_i},
\]
with $\zeta_i\equiv \partial \ln Z_i /\partial M.$
When $M, N$ is very large, the summation in $Z_i$ runs over many states and it's insensitive to which agent is ignored \cite{statistical_mechanics}. Therefore $\zeta_i=-E_0$ is a parameter independent of $i$, and  
\begin{equation}
Z=\frac{M!}{g!}Z_i(M)\sum_{m_i=0}^gC_g^{m_i}\e^{-(E_i-E_0)m_i}
=\frac{M!}{g!}Z_i(M)(1+\e^{-(E_i-E_0)})^g.
\end{equation}
from which we can obtain
\begin{equation}
    p_i=-\frac{\partial\ln Z}{\partial(E_i-E_0)}=\frac{g}{\e^{E(x_i)-E_0}+1}.
\end{equation}
$E_0$ can be determined by the constraint $\sum_{i=0}^Np_i=M$.

Therefore, in the thermodynamic limit 
\begin{equation}\label{eq:px_g}
    p(x)=\frac{g}{\e^{E(x)-E_0}+1}=\frac{g}{\e^{\alpha(x-x_0)}+1},
\end{equation}
where we substitute $E(x)=\alpha x$ and $x_0\equiv E_0/\alpha$. $x_0$ (and $E_0$) can be determined from $\int p(x)\rho(x,t)=m_t$
with $m_t\equiv M/N$. 

We consider a travelling wave solution $\rho(x,t)=\rho(u)$ where $u\equiv x-vt$ and $v=ke_pm_t$ is the drift speed which we already obtained. 
The steady traveling wave solution $\rho_s(u)$ is determined by
\begin{equation}\label{mean-field}
0=\frac{\partial \rho_s}{\partial t}=
-kg\frac{\partial}{\partial u}\left(
e_p(\frac1{\e^{\alpha (u-u_0)}+1}-A_t)\rho_s-
\frac{\partial}{\partial u}
\frac1{\e^{\alpha (u-u_0)}+1}\rho_s
\right),
\end{equation}
with $A_t\equiv m_t/g$, from which we can directly integrate over $u$ and obtain
\begin{equation}\label{eq:rho}
\rho_{s,g}(u)=\frac1Z(\e^{\alpha(u-u_0)}+1)\exp[e_p(1-A_t)(u-u_0)-\frac{A_te_p}{\alpha}e^{\alpha(u-u_0)}],
\end{equation}
where $Z$ is a normalization constant.
In this expression, parameter $u_0$ only determines the position of the wave packet but does not affect the waveform, so it's self-consistent to arbitrarily choose $u_0$ from any real number. For simplicity we choose $u=0$ in main text. 
In this case, by calculating the derivative we can determine that the maximum position of $\rho_s(u)$ (denoted by $u_m$) is
\begin{equation}
u_m=\frac1\alpha\ln\left(
\frac{e_p(1-2A_t)+\alpha+\sqrt{(e_p+\alpha)^2-4e_p\alpha A_t}}{2A_te_p}
\right).
\end{equation}

We can determine the probability flux by 
\[
J(x,t)\equiv ke_pp(x)\rho-k\frac{\partial}{\partial x}p(x)\rho
=v\rho(x,t),
\]
so that the total dissipation rate for this steady traveling wave solution is
\begin{align}
\dot{W}&=\int\frac{J^2}{kp(x)\rho_s}\d x
=\frac{v^2}{kg}
\int_{-\infty}^{\infty}\rho_s(1+\e^{\alpha u})\d u \notag\\
&=ve_pA_t(1+\frac{1-A_t}{A_t}\frac{e_p+\alpha}{e_p})
=v[e_p+(1-A_t)\alpha],
\label{eq:diss_exp}
\end{align}
where the integral can be calculated by using $\Gamma$-function. 

In \eqref{eq:rho}, when $\alpha\to\infty$, the term:
\begin{equation}
\int\frac1Z \e^{\alpha(u-u_0)}\times\exp[e_p(1-A_t)(u-u_0)-\frac{A_te_p}{\alpha}e^{\alpha(u-u_0)}]
=1-A_t,
\end{equation}
which keeps finite no matter how large $\alpha$ is. Meanwhile, this term is almost 0 everywhere when $\alpha\to\infty$, except for a singularity point at $x=0$ for very large $\alpha$, which is characteristic of a $\delta$-function. It's also easy to check that the rest terms in \eqref{eq:rho} converge to an exponential distribution. Taking these two parts together, we have the expression of $\rho_s$ in large $\alpha$ limit (denoted by $\rho_{s,\infty}$)
\begin{equation}
\rho_{s,\infty}(u)=C_1\delta(u)+C_2\rho_e(u)H(-u), 
\end{equation}
where $H$ is the Heaviside step function and $\rho_e(u)=\lambda e^{\lambda u}$ with $\lambda=e_p(1-A_t)$. $C_1=1-A_t$ and $C_2=A_t$ are the weights of the two distributions.


\section{The energy-speed-accuracy relation for pairwise-interacting model}
In this section we briefly discuss the energy-speed-accuracy (ESA) relation in the direct pairwise-interaction model, whose synchronization dependence on free energy cost is studied in our previous work \cite{sync_energy}.

In that work, we defined an order parameter $r$ for a group of oscillators as:
\begin{equation}
    r\equiv \langle \e^{i \theta} \rangle,
\end{equation}
where $\theta$ is the relative phase. 
In strong-synchronization limit, $\theta$ is symmetrically distributed near 0 and $r$ approaches 1. This allows we to expand 
$r\approx \langle 1-\frac12\theta^2+\ldots \rangle$, and obtain the relation between the variance of $\theta$ (denoted by $\sigma_\theta^2$) and $r$:
\begin{equation}
    \sigma^2_\theta\equiv\langle \theta ^2\rangle\approx 2(1-r).
\end{equation}

Meanwhile, in the Eq.15 of Ref.\cite{sync_energy}, we derived that the maximum $r$ with a given total energy budget is:
\begin{equation}
    r_{max}(W_T)=1-\frac\pi{e_g(W_T-W_{0,T})},
\end{equation}
where $W_T$ and $W_{0,T}$ are the total energy and the energy to purely drive individual oscillators respectively in a period $T$, $e_g\equiv W_0/(2\pi)$. With this relation we can easily derive the relation between the maximum of $\sigma^2_\theta$ and energy cost:
\begin{equation}\label{eq:compare:PI}
    \sigma^2_{\theta,max}=2(1-r_{max})
    =\frac{\omega^2}{\dot{W}_{0}\dot{W}_{s}},
\end{equation}
where $\omega\equiv 2\pi/T$ is the angular velocity, 
$\dot{W}_{0}\equiv W_0/T$ is the dissipation rate for driving individual oscillators, and
$\dot{W}_{s}\equiv W_{T}/T-\dot{W}_{0}$ is the dissipation for synchronization. 
This relation shows the same quantitative ESA trend with Eq.12 in main text, which has the form:
\begin{equation}\label{eq:compare:ARS}
    \sigma^2
    =\frac{v^2}{\dot{W}_{0}\dot{W}_{s}}.
\end{equation}

\begin{figure}
\begin{minipage}{\linewidth}
\centering
\includegraphics[width=0.5\linewidth]{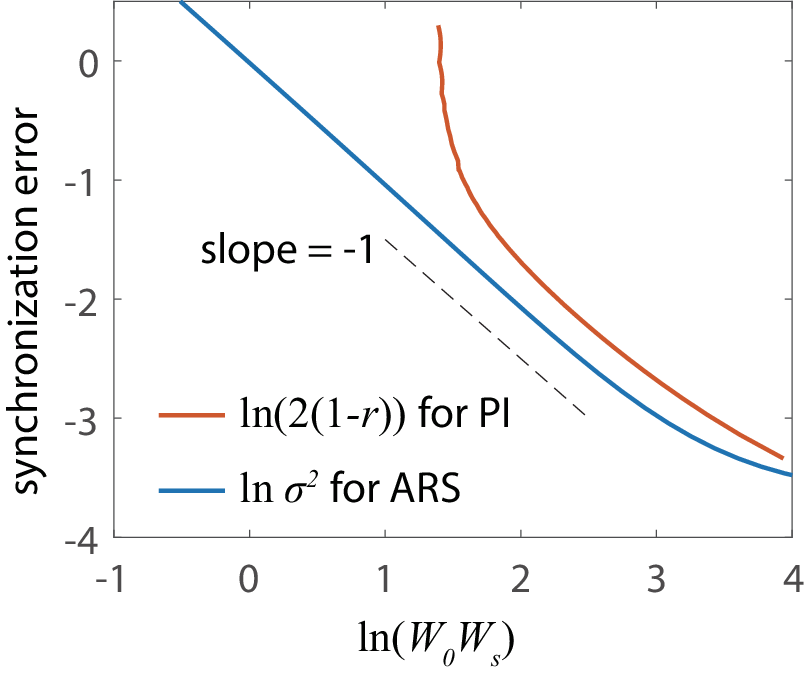}
\end{minipage}
\caption{
Numerical calculation for how the synchronization error decreases with energy dissipation. In general, the error is suppressed in both mechanisms. Although there are differences between the two curves, quantitatively they both show a inverse-dependence regime.
}
\label{fig:compare}
\end{figure}

Fig.S\ref{fig:compare} illustrates the numerical calculation for how the synchronization error decreases with energy dissipation. The synchronization error is defined as $\sigma^2$ for ARS mechanism and $1-r$ for PI mechanism.
In general, the error is suppressed with increasing energy dissipation in both mechanisms. Although there are differences between the two curves, quantitatively they both show a common inverse-dependence regime, as shown by \eqsref{eq:compare:PI} and (\ref{eq:compare:ARS}).

\section{Comparison with TUR}
In this section, we briefly discuss the application of thermodynamic uncertainty relation (TUR) \cite{TUR} in this problem. 

To relate our problem to TUR, we need to choose an observable whose current has a clear physical meaning. 
One choice is to track how long the agent travels within a time interval $T$, as did in Ref.\cite{lee2018thermodynamic}, denoted by $l_T\equiv \int j_{x,T}\d x$, with $j_{x,T}$ the net number of transitions (events) from state $x$ to $x+\d x$ within $T$. From Eq.4 in main text, it's average value is
\begin{equation}
    \langle l_T \rangle =
    \int_{-\infty}^{+\infty}
     \left[
    \int_0^T\left(
    ke_pp\rho-\frac{\partial p\rho}{\partial x}
    \right)\d t
    \right]\d x=vT,
\end{equation}
where $v=ke_pm_t$ is exactly the mean velocity. 

\begin{figure}
\begin{minipage}{\linewidth}
\centering
\includegraphics[width=0.7\linewidth]{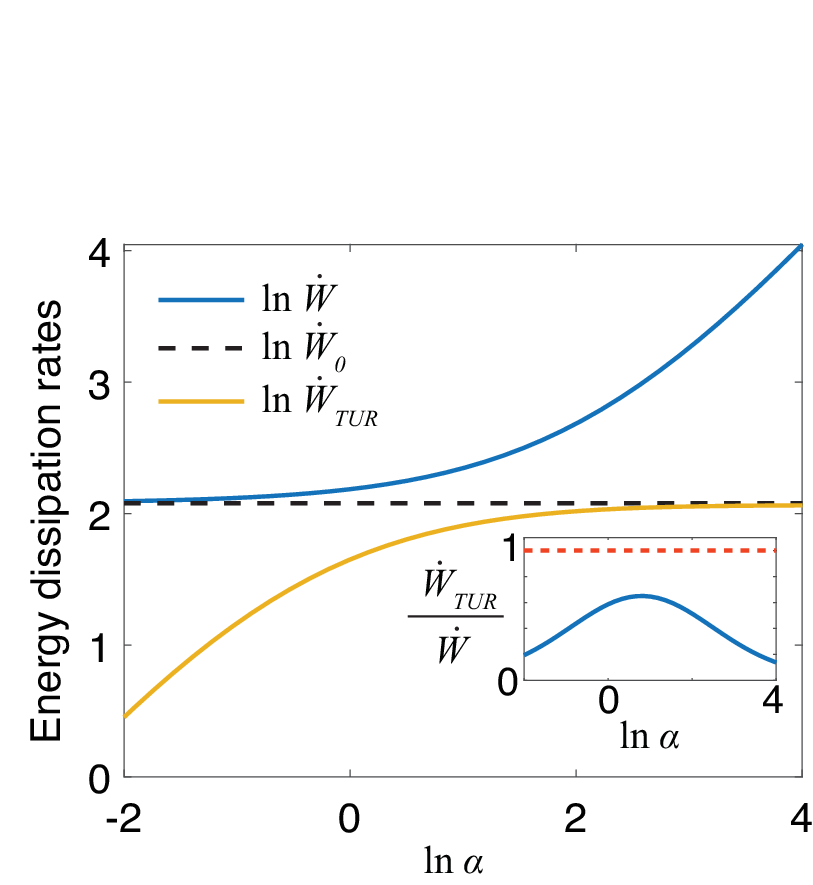}
\end{minipage}
\caption{
Comparison between $\dot{W}, \dot{W}_0$, and $\dot{W}_{TUR}$. It's clear that $\dot{W}>\dot{W}_0>\dot{W}_{TUR}$, and $\dot{W}_{TUR}$ cannot actually reach $\dot{W}$. Inset: the ratio of $\dot{W}_{TUR}/\dot{W}$, which is always smaller than 1.
}
\label{fig:rho(u)}
\end{figure}

The variance $\sigma_l^2$ is made up of two parts $\sigma_l^2=\sigma_0^2+\sigma^2$, where $\sigma^2$ is the variance of $u$, and $\sigma_0^2$ is the variance for mean position of all $N$ agents, which is thus averaged over all the agents in the systems: 
\begin{equation}
    \sigma_0^2=\frac1N
    \int_{-\infty}^{+\infty}
    \int_0^T
    2kp(x)\rho(x)\d t\d x
    =\frac{T}{N}\int_{-\infty}^{+\infty}2kp(x)\rho(x)\d x,
\end{equation}
where we have used the condition that $\rho(x,t)$ is a traveling wave.
In this way, we can derive the expression of $\sigma_l^2$:
\begin{equation}
    \sigma_l^2=\frac{T}{N}\int_{-\infty}^{+\infty}2kp(x)\rho(x)\d x+\sigma^2
    =\frac{2Tv}{Ne_p}+\sigma^2,
\end{equation}
where we applied $\int p(x)\rho(x)=m_t$ and $v\equiv ke_pm_t$.
The lower bound for dissipation rate is thus set by TUR:
\begin{equation}
\dot{W}_{ TUR}\equiv \frac{2\langle l_{T} \rangle^2}{T\sigma_{l}^2}
   =\frac{\dot{W}_0}{1+\frac{e_p\sigma^2 N}{2vT}},
\end{equation}
where we used $\dot{W}_0=ve_p$.
$\dot{W}_{TUR}$ is no larger than $\dot{W}_0$ and the actual dissipation rate $\dot{W}=v(e_p+(1-A_t)\alpha)=\dot{W}_0+\dot{W}_s$. The energy due to synchronization $\dot{W}_s$, which is physically more important in our problem, is abscent in TUR.
$\dot{W}_0$ and $\dot{W}_{TUR}$ are equal only when $\rho(x,t)$ is uniform indicating no synchronization at all. Hence $\dot{W}_{TUR}$ is always smaller than $\dot{W}$ as long as there is non-zero $\alpha$ and there is synchronization among the agents, as illustrated in Fig.S\ref{fig:rho(u)}.

\section{The effect of nonlinear E(x)}
Although our previous results use the linear form of $E(x)=\alpha x$, the result can be extended to understand the behavior of arbitrary $E(x)$. 
If $E(x)$ is an arbitrary increasing function of $x$ in \eqref{eq:px_g}, there is no general steady traveling wave solution for \eqref{eq:rho} that keeps the waveform invariant. 
But if $E(x)$ does not vary too sharply, i.e., the derivative of $E(x)$ changes slowly, 
the waveform of $\rho_s(x,t)$ will changes adiabatically. 
In this way, \eqref{eq:rho} serves as a zeroth-order approximation, with 
\begin{equation}
    \alpha=|\frac{\d E(x)}{\d x}|_{x=x_m},
\end{equation}
now defined as the derivative of $E(x)$ at $x_m=\arg\max_x\rho(x,t)$ where $\rho(x,t)$ reaches its maximum. 
Similarly, the ESA relation should also hold approximately.

\begin{figure}
\begin{minipage}{\linewidth}
\centering
\includegraphics[width=0.95\linewidth]{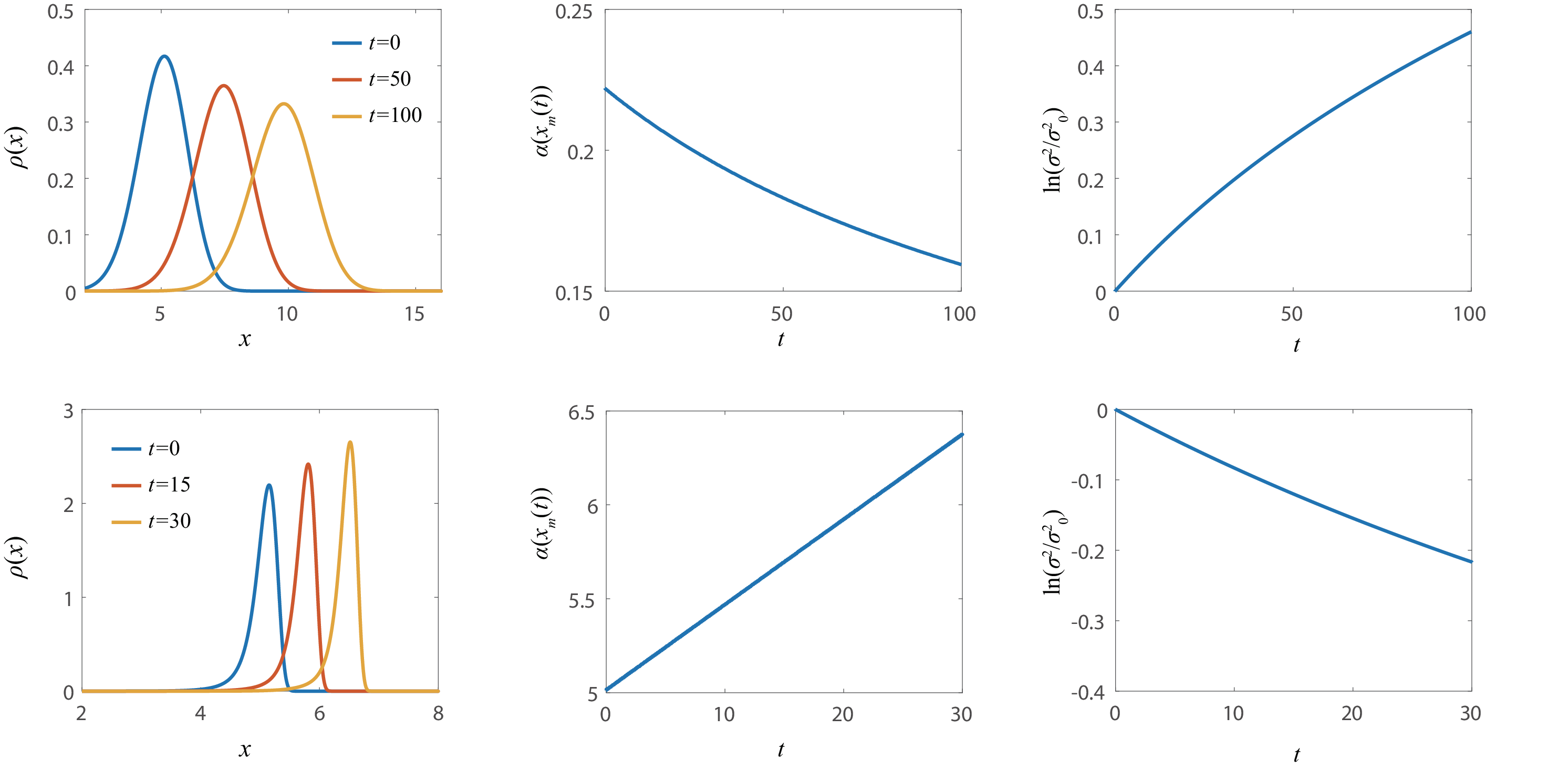}
\end{minipage}
\caption{
Numerical behaviors of how $\rho(x)$ evolves with nonlinear $E(x)$. 
Simulation is done under the parameter of finite $A_t=0.16$. Upper panels: $E(x)=\alpha_0x^{1/2}$. Lower panels: $E(x)=\alpha_0x^{2}$. $\sigma_0$ is the $\sigma$ at $t=0$. 
}
\label{fig:sigma_t}
\end{figure}

We can also obtain the necessary condition for $E(x)$ to generate effective synchronization. In addition to the requirement for $E(x)$ to be monotonically increasing, its derivative must remain non-zero at all times to prevent $\sigma$ from diverging. This implies that $E(x)$ should grow no slower than linearly as $x\to\infty$, and consequently, $p(x)$ must decay faster than exponentially as $x\to\infty$. Fig.S\ref{fig:sigma_t} illustrates how $\sigma^2$ decays (diverges) with time for $E(x)$ faster (slower) than linearly, and shows the change in corresponding $\alpha(x_m(t))$, based on simulations of the Fokker-Planck equation \eqref{mean-field} with corresponding $p(x)$. This confirms this conclusion.

\section{The effect of ignoring free activators}
In this work, we mostly focus on the regime where $\e^{E(x)}\ll1$, i.e. for $x\ll x_0$ where $E(x_0)=0$, and ignore free activators. In this section, we briefly discuss the case when the center of the traveling wave approaches $x_0$ and how this approximation should be fixed. 

Mathematically, we can introduce a ratio 
$R$
defined as the fraction of the free activators. 
In this way, 
the conservation of the activators should be modified as
\begin{equation}
   \int p(x)\rho(x,t)=gA_t(1-R)\equiv gA_{eff},
\end{equation}
which is not a strict constant now, and the drift speed should be modified as $v=ke_pA_{eff}$ is also not a constant in principle.
When $x$ is small,
$A_{eff}$ is quite close to $A_t$ and can be regarded as a constant. When $x$ approaches $x_0$, $R$ dramatically increases and $A_{eff}$ quickly drops, which prevents the agents from moving further. 
Specifically, since $p(x)\sim \e^{-E(x)}$ for large $x$, for large $R$ we have 
\begin{equation}
    1-R=\frac{\int p(x)\rho(x,t)}{M_0/N+\int p(x)\rho(x,t)}
    \sim\e^{-\alpha(x_m-x_0)},
\end{equation}
where $M_0$ denotes the number of free activators, and $x_m$ is the maximum point of $\rho_s(x,t)$. Therefore, $A_{eff}$ drops exponentially with $(x_m-x_0)$, and so will the average velocity $v=ke_pA_{eff}$. 
As $A_{eff}$ drops, $\sigma$ will also decrease just as if $A_t$ decreases in \eqref{eq:rho}.


\begin{figure}
\begin{minipage}{\linewidth}
\centering
\includegraphics[trim=0 150 0 150, clip, width=0.9\linewidth]{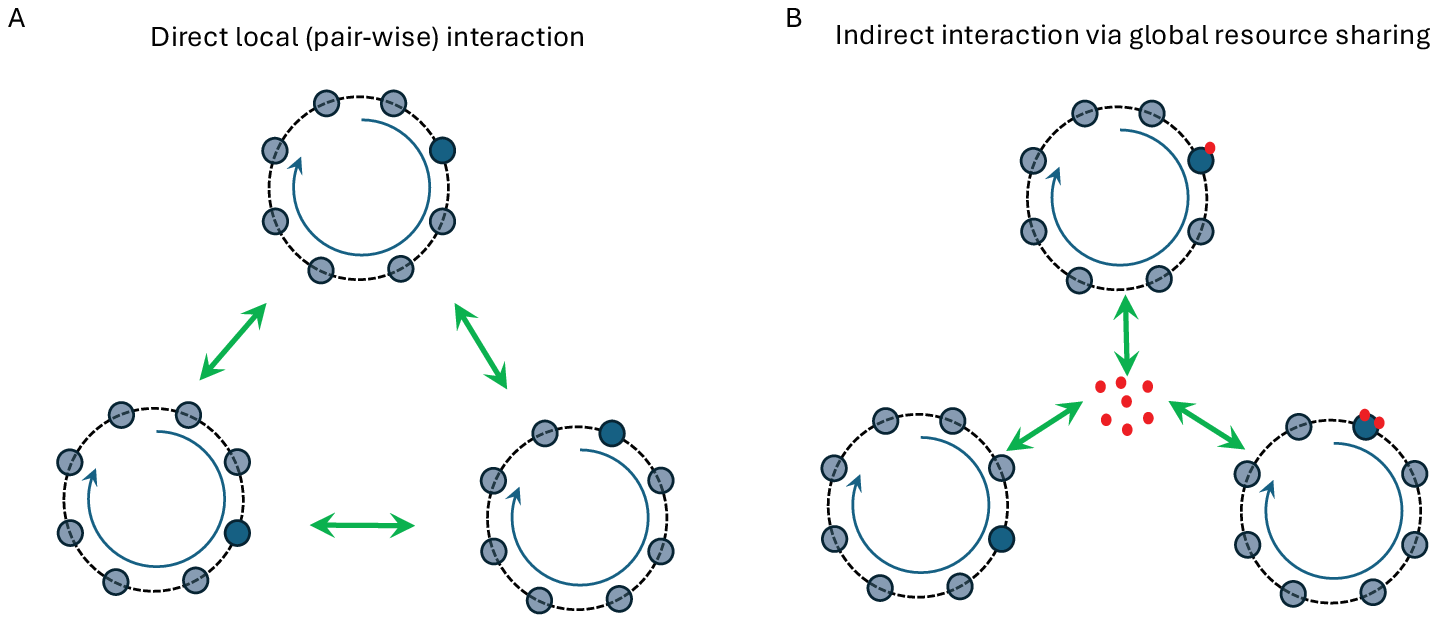}
\end{minipage}
\caption{
Illustration of the two synchronization mechanisms. Each ring represents a oscillator with its possible states represented by the blue circles. The darker blue circle represents the current state of the oscillator. (A) The local pair-wise interaction (PI) mechanism where the pair-wise interactions (green-arrowed lines) tend to decrease the difference between the two interacting oscillators. (B) The global indirection interaction mechanism where the oscillators interact through sharing resources (red dots), which is needed for advancing the phase of the oscillation. To achieve synchronization, the altruistic resource sharing (ARS) mechanism allows the lagger (the oscillator on the bottom right) to gain more resources than the oscillators at the top or the bottom left that are more advances.}
\label{fig:illus}
\end{figure}

\bibliography{Sync_diff_ref.bib}